# Comprehensive first-principles insights into the physical properties of intermetallic $Zr_3Ir$: a noncentrosymmetric superconductor

Razu Ahmed[1,2], Md. Sajidul Islam[1,2], M. M. Hossain[2,3], M. A. Ali[2,3], M. M. Uddin[2,3], S. H. Naqib[1,2,*]
[1]Department of Physics, University of Rajshahi, Rajshahi 6205, Bangladesh
[2]Advanced Computational Materials Research Laboratory, Department of Physics, Chittagong University of Engineering and Technology (CUET), Chattogram-4349, Bangladesh
[3]Department of Physics, Chittagong University of Engineering and Technology (CUET), Chattogram-4349, Bangladesh
*Corresponding author, Email: salehnaqib@yahoo.com

## Abstract

We have looked into the structural, mechanical, optoelectronic, superconducting state and thermophysical aspects of intermetallic compound $Zr_3Ir$ using the density functional theory (DFT). Many of the physical properties, including direction dependent mechanical properties, Vickers hardness, optical properties, chemical bonding nature, and charge density distributions, are being investigated for the first time. According to this study, $Zr_3Ir$ exhibits ductile features, high machinability, significant metallic bonding, a low Vickers hardness with low Debye temperature, and a modest level of elastic anisotropy. The mechanical and dynamical stabilities of $Zr_3Ir$ have been confirmed. The metallic nature of $Zr_3Ir$ is seen in the electronic band structures with a high electronic energy density of states at the Fermi level. The bonding nature has been explored by the charge density maping and bond population analysis.The tetragonal $Zr_3Ir$ shows a remarkable electronic stability, as confirmed by the presence of a pseudogap in the electronic energy density of states at the Fermi level between the bonding and antibonding states. Optical parameters show very good agreement with the electronic properties. The reflectivity spectra reveal that $Zr_3Ir$ is a good reflector in the infrared and near-visible regions. $Zr_3Ir$ is an excellent ultra-violet (UV) radiation absorber. High refractive index at visible photon energies indicates that $Zr_3Ir$ could be used to improve the visual aspects of electronic displays. All the optical constants exhibit a moderate degree of anisotropy. $Zr_3Ir$ has a moderate melting point, high damage tolerance, and very low minimum thermal conductivity. The thermomechanical characteristics of $Zr_3Ir$ reveal that it is a potential thermal barrier coating material. The superconducting state parameters of $Zr_3Ir$ are also explored.

**Keywords:** DFT calculations; Elastic properties; Thermophysical properties; Optoelectronic properties; Superconductivity

## 1. Introduction

The iridium-based refractory super alloys have drawn attention as potential high-temperature structural materials due to their outstanding combination of physical properties [1]. In recent years, Zr-Ir super alloys have been receiving significant attention. Because of high melting temperature and stability, the intermetallic compound $ZrIr_3$ exhibits very high strength, even at elevated temperatures and has great potential to be used as a structural material in aircrafts engineering [2]. In 1996, a novel type of refractory Ir-based super alloys was discovered for ultra-high temperature applications (1800°C) [3]. Kuprina & Kuruyachaya investigated the Zr-Ir phase diagram and found that binary compounds existed at five distinct compositions [4]. Only two phases were confirmed on the Zr-rich side of the phase diagram, namely, $Zr_2Ir$ with $CuAl_2$-type structure (McCarthy, 1971) and $Zr_3Ir$. Raman and Schubert studied the lattice characteristics of $Zr_3Ir$ and determined it as $\alpha$-$V_3S$ structure and identified it as being isotypic with $\beta$-U [4]. Moreover, the compound, $Zr_3Ir$ exhibits s-wave superconductivity with a transition temperature of $T_c$ = 2.3 K, confirmed by the magnetization, electronic specific heat, and muon spin rotation studies [5].

In this study, the compound, $Zr_3Ir$ is investigated using first-principles calculations based on the density functional theory. Some earlier experimental and theoretical results on the chosen material were documented in Refs.[2–5]. In those studies, structural, mechanical, electronic (band structure and density of states), and some of the thermodynamic properties have been studied. Several significant physical characteristic such as optical properties, chemical bonding nature have not yet been studied. Additionally, there hasn't been any report on the mechanical anisotropy characterized by the directional dependence of the Young's modulus, shear modulus, linear compressibility, and Poisson's ratio of $Zr_3Ir$. The hardness and machinability index of this compound remain unexplored.

We have looked at the physical properties of $Zr_3Ir$ in order to fully understand its structural, elastic, bonding, electronic, optical, thermo-physical, and superconducting state properties. The results obtained are compared with those found in previous studies where available. The electronic properties, such as electronic band structure, density of states, Fermi surface topology and charge density distribution are connected to charge transport, optical, and electronic thermal processes. Mulliken and Hirshfeld bond population analyses are used to provide an explanation for the bonding nature of the compound under consideration. The type of chemical bonding in solids affects their mechanical properties. In addition, knowledge of bond hardness, fracture toughness, and brittleness index is helpful for designing any component of a structure or device, and these are also computed in this study.

The rest of the paper is organized in the following manner: In Section 2, we have discussed the computational scheme. Section 3 discloses the computational results and analyses. Finally, the major findings of this study are discussed and summarized in Section 4.

## 2. Computational scheme

In density functional theory (DFT), the ground state of a system is obtained from the solution of the Kohn-Sham equation [6]. DFT based on the plane wave pseudopotential approach as implemented in the CAmbridge Serial Total Energy Package (CASTEP) code [7], has been used to investigate the aforementioned properties of $Zr_3Ir$. We have used the generalized gradient approximation (GGA) with the Perdew–Burke–Ernzerhof (PBE) exchange-correlation functionals. Considering the experimental structural parameters, GGA (PBE) [8] provides the best results for the ground-state crystal geometry. Ultrasoft Vanderbilt-type pseudopotentials have been employed to calculate the interaction between the valence electrons and ion cores [9]. The use of ultra-soft pseudopotential saves substantial computational time without affecting the accuracy appreciably. The valence electron configurations of Zr and Ir are taken as [$4s^2 4p^6 4d^2 5s^2$] and [$5d^7 6s^2$], respectively. Monkhorst-Pack scheme with a mesh size of 3 × 3 × 6 *k*-points for $Zr_3Ir$ has been selected and a cut-off energy of 400 eV is set. To obtain a smooth and correct Fermi surface of $Zr_3Ir$ compound, a 10 × 10 × 18 *k*-points mesh has been used. The BFGS (Broyden-Fletcher- Goldfarb-Shanno) algorithm [10] has been employed to optimize the crystal structure. The X-ray diffraction (XRD) pattern and electronic structure of $Zr_3Ir$ have been visualized using the VESTA program [11]. The structure has been relaxed with a convergence limit of $5\times10^{-6}$ eV-atom$^{-1}$ for energy, 0.01 eV Å$^{-1}$ for maximum force, 0.02 GPa for maximum stress, and $5\times10^{-4}$ Å for maximum atomic displacement. All first-principles calculations have been carried out at ground state, with default temperature and pressure of 0 GPa and 0 K, respectively.

The elastic constants, $C_{ij}$, are calculated using the stress-strain technique [12]. Only six independent elastic constants ($C_{11}$, $C_{33}$, $C_{44}$, $C_{66}$, $C_{12}$, and $C_{13}$) exist for tetragonal crystals. Using the Voigt-Reuss-Hill (VRH) approximation [13,14], all the other polycrystalline elastic parameters, including the bulk modulus (*B*), shear modulus (*G*), and Young modulus (*Y*) can be computed from the values of the computed $C_{ij}$. The electronic band structure, total density of states (TDOS), and atom-resolved partial density of states (PDOS) are determined using the optimized crystal structures of $Zr_3Ir$.

The complex dielectric function, $\varepsilon(\omega) = \varepsilon_1(\omega) + i\varepsilon_2(\omega)$, can be used to obtain all the energy/frequency dependent optical constants. The real part of the dielectric function is obtained from the imaginary part employing the Kramers-Kronig transformation equation. The momentum representation of the matrix elements of the photon-induced transition between the occupied and unoccupied electronic states is used to obtain the imaginary part, $\varepsilon_2(\omega)$, of the complex dielectric function. The CASTEP contained expression giving the imaginary part of the dielectric function is shown below:

$$\varepsilon_2(\omega) = \frac{2e^2\pi}{\Omega\varepsilon_0}\sum_{k,v,c}|\langle\psi_k^c|\hat{u}.\,\vec{r}|\psi_k^v\rangle|^2\,\delta(E_k^c - E_k^v - E) \qquad (1)$$

where, Ω denotes the cell volume, *ω* is the angular frequency of incident electromagnetic wave (EMW), $\hat{u}$ is a unit vector giving the polarization direction of the electric field, *e* is the electronic charge, $\psi_k^c$ and $\psi_k^v$ are the conduction and valence band eigenstates (at a fixed wave vector *k*). The conservation of energy and momentum is forced by the delta function in Equation (1). All the other important optical parameters are calculated from the real and imaginary parts of the dielectric function $\varepsilon(\omega)$, using well-established interrelations [15,16].

The real, $n(\omega)$, and imaginary, $k(\omega)$, parts of the complex refractive index can be computed by using the following relations:

$$n(\omega) = \frac{1}{\sqrt{2}}[\{\varepsilon_1(\omega)^2 + \varepsilon_2(\omega)^2\}^{1/2} + \varepsilon_1(\omega)]^{1/2} \qquad (2)$$

$$k(\omega) = \frac{1}{\sqrt{2}}[\{\varepsilon_1(\omega)^2 + \varepsilon_2(\omega)^2\}^{1/2} - \varepsilon_1(\omega)]^{1/2} \qquad (3)$$

Again, the reflectivity, $R(\omega)$, can be calculated by using the complex refractive index components:

$$R(\omega) = \left|\frac{\tilde{n}-1}{\tilde{n}+1}\right| = \frac{(n-1)^2+k^2}{(n+1)^2+k^2} \quad (4)$$

Furthermore, the absorption coefficient, $\alpha(\omega)$, the optical conductivity, $\sigma(\omega)$, and the energy loss function, $L(\omega)$, can be determined from the following equations:

$$\alpha(\omega) = \frac{4\pi k(\omega)}{\lambda} \quad (5)$$

$$\sigma(\omega) = \frac{2W_{cv}\hbar\omega}{E^2} \quad (6)$$

$$L(\omega) = Im\left(-\frac{1}{\varepsilon(\omega)}\right) \quad (7)$$

In Equation (6), $W_{cv}$ is the transition probability per unit time.

The Mulliken bond population study [17] is a popular framework for comprehending a material's bonding characteristics. A projection of the plane-wave states onto a linear combination of atomic orbital (LCAO) basis sets [18,19] is used for $Zr_3Ir$.

The Mulliken bond population analysis can be implemented using the Mulliken density operator as follows:

$$P^{M}_{\mu\,v}(g) = \sum_{g'}\sum_{v'} P_{\mu v'}(g')S_{v'v}(g-g') = L^{-1}\sum_{k} e^{-ikg}(P_k S_k)_{\mu v'} \quad (8)$$

and the net charge on an atomic species $A$ is defined as:

$$Q_A = Z_A - \sum_{\mu\in A} P^{m}_{\mu\mu}(o) \quad (9)$$

where $Z_A$ represents the charge of the nucleus.

## 3. Results and discussions

### *3.1. Structural properties and enthalpy of formation*

The compound, $Zr_3Ir$ belongs to a non-centrosymmetric tetragonal $\alpha$-$V_3S$ crystal structure with space group 121 (*I-42m*). The schematic crystal structure of $Zr_3Ir$ is displayed in Figure 1. Eight formula units and 32 atoms make up the unit cell of $Zr_3Ir$ of which 24 are zirconium atoms and 8 are iridium atoms. The Zr and Ir atoms occupy the following Wyckoff positions in the unit cell [4], Zr atoms: (0.0952, 0.0952, 0.2610); (0.2940, 0.2940, 0.2536); (0.3544, 0, 0.5) and Ir atoms: (0.2919, 0, 0). In Table 1, the structural properties of $Zr_3Ir$ along with previously obtained experimental and theoretical lattice parameters are enlisted. The calculated lattice parameters match the experimental findings very well.

In order to investigate the phase stability for $Zr_3Ir$, the formation enthalpy has been calculated and the outcome is enclosed in Table 1. The formation enthalpy of $Zr_3Ir$ has been calculated using the following equation [20]:

$$\Delta H_f(Zr_3Ir) = \frac{E_{Total}(Zr_3Ir) - xE(Zr) - yE(Ir)}{x + y} \qquad (10)$$

where, $E_{Total}(Zr_3Ir)$ is the total enthalpy of $Zr_3Ir$ compound; $E(Zr)$ is the total enthalpy of the Zr atom and $E(Ir)$ is the total enthalpy of the Ir atom in the solid state. We have $x$ = 24 for Zr and $y$ = 8 for Ir atoms in the unit cell.

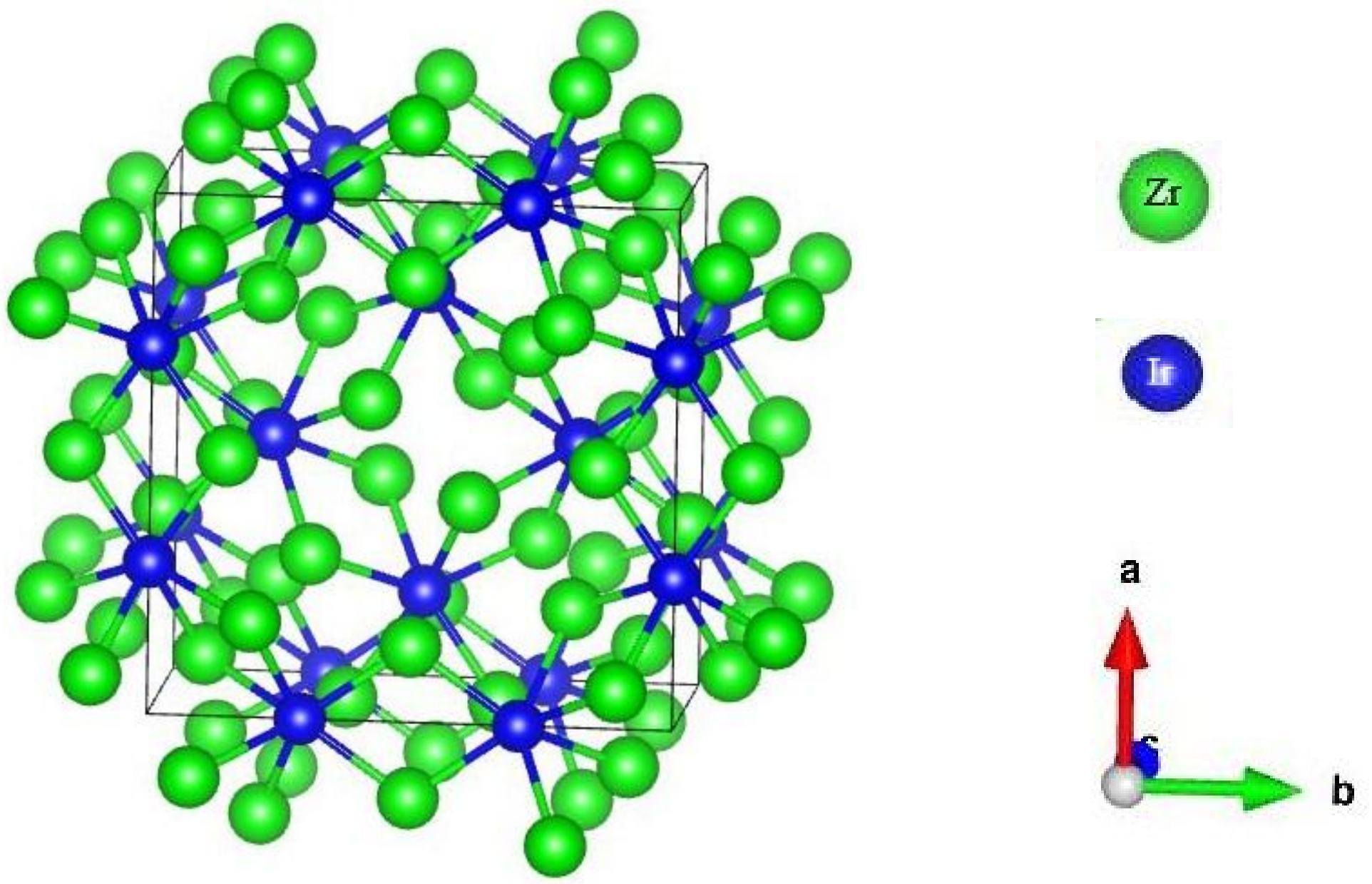

**Figure 1.** Schematic crystal structure of $Zr_3Ir$. The crystallographic axes are shown on the right.

**Table 1**.

Calculated and experimental lattice constants $a$ and $c$ (Å), equilibrium cell volume $V_o$ (Å$^3$), and formation enthalpy (KJ/mol) of $Zr_3Ir$.

| Compound | $a = b$ | $c$ | $V_o$ | Formation enthalpy | Ref. |
|---|---|---|---|---|---|
| $Zr_3Ir$ | 10.790 | 5.790 | 664.703 | -820.04 | This work |
| | 10.788 | 5.602 | 651.966 | --- | [5]$^{Expt.}$ |
| | 10.811 | 5.708 | 667.138* | --- | [3]$^{Theo.}$ |

*Calculated from the lattice constants.

### *3.2. Elastic properties*

Most of the solid-state properties of crystalline materials, such as brittleness, ductility, stiffness, structural stability, normal modes of oscillations, elastic wave propagation, and anisotropy, must be understood in order to predict how a material will respond when external stress is applied. The tetragonal compound $Zr_3Ir$ has six independent elastic constants ($C_{11}$, $C_{33}$, $C_{44}$, $C_{66}$, $C_{12}$, and $C_{13}$). The calculated elastic constants are listed in Table 2. According to Born-Huang conditions, a tetragonal crystal system satisfy the following criteria for mechanical stability [21,22]:

$$C_{11} > 0, C_{33} > 0, C_{44} > 0, C_{66} > 0, (C_{11} - C_{12}) > 0, (C_{11} + C_{33} - 2C_{13}) > 0,$$
$$\{2(C_{11} + C_{12}) + C_{33} + 4C_{13}\} \qquad (11)$$

All the elastic constants of $Zr_3Ir$ are positive and satisfy the above mentioned criteria, which implies that $Zr_3Ir$ is mechanically stable. Here, it is seen that $C_{33} > C_{11}$, it indicates that the bonding strength along [001] direction is stronger than that along [100] direction. The relation $C_{44} > C_{66}$ for $Zr_3Ir$ predicts that the deformation along [100] (001) shear plane is expected to be more difficult than along the [100] (010) shear plane. The tetragonal shear modulus, [$C' = (C_{11} - C_{12})/2$] of a crystal is the measure of crystal's stiffness and its positive value suggests the dynamical stability of the solid. Thus, the compound under investigation is expected to be dynamically stable. Table 2 also shows the results from a prior theoretical work, showing a very high level of agreement with our values.

**Table 2**

Calculated elastic constants, $C_{ij}$ (GPa) and tetragonal shear modulus, $C'$ (GPa) of $Zr_3Ir$.

| Compound | $C_{11}$ | $C_{33}$ | $C_{44}$ | $C_{66}$ | $C_{12}$ | $C_{13}$ | $C'$ | Ref. |
|---|---|---|---|---|---|---|---|---|
| $Zr_3Ir$ | 161.04 | 177.33 | 35.60 | 21.55 | 137.05 | 98.00 | 12.00 | This work |
| | 166.00 | 172.70 | 36.60 | 22.80 | 136.70 | 96.40 | --- | [3] |

The Hill approximated values of bulk modulus ($B_H$) and shear modulus ($G_H$) (using the Voigt-Reuss-Hill (VRH) method), Young's modulus ($Y$), Poisson's ratio ($\sigma$), Cauchy pressure, ($C''$), machinability index ($\mu_M$), Kleinman parameter ($\zeta$) of $Zr_3Ir$ have been computed using the following standard formulae [23–29]:

$$B_H = \frac{B_V + B_R}{2} \qquad (12)$$

$$G_H = \frac{G_V + G_R}{2} \qquad (13)$$

$$Y = \frac{9BG}{(3B + G)} \qquad (14)$$

$$\sigma = \frac{(3B - 2G)}{2(3B + G)} \quad (15)$$

$$C'' = (C_{12} - C_{44}) \quad (16)$$

$$\mu_M = \frac{B}{C_{44}} \quad (17)$$

$$\zeta = \frac{C_{11} + 8C_{12}}{7C_{11} + C_{12}} \quad (18)$$

The computed values are presented in Table 3. From Table 3, we see that $B > G$, which implies that the shearing stress should determine the mechanical stability of $Zr_3Ir$ and the mechanical failure mechanism of $Zr_3Ir$ will also be determined by the shape deforming stress, rather than the volume changing stress. The value of $Y$ is a measurement of a material's stiffness (resistance to length change) [30,31]. A material is more stiffer the higher its Young's modulus. Also, the higher the value of Young's modulus, the more covalent a material is [32]. Moreover, the lattice thermal conductivity ($K_L$) of solids is roughly related to its Young’s modulus as follows: $K_L \sim \sqrt{Y}$ [33].

Some of the indicators that classify whether a material is brittle or ductile include the Pugh’s ratio ($B/G$), Poisson’s ratio ($\sigma$), and Cauchy pressure ($C''$). The critical value of Pugh’s ratio is 1.75 [34–36], Poisson’s ratio is 0.26 [37,38], and Cauchy pressure is zero [27,29]. A material is ductile (brittle), if the value of $B/G$ is greater (less) than 1.75, $\sigma$ is higher (lower) than 0.26, and $C''$ is positive (negative) [39]. The calculated values of these parameters, which are shown in Table 3, suggest that $Zr_3Ir$ is ductile.

It is known that the Poisson’s ratio ($\sigma$) of a crystal, which is defined as the absolute value of the ratio of transverse strain to longitudinal strain, can provide certain details about a crystal's bonding force and anisotropy [34,40]. The value of $\sigma$ is between 0.25 and 0.50 for central-force dominated solids [40]. For $Zr_3Ir$, the value of $\sigma$ is approximately 0.40, indicating that the interatomic force is the central force type. Also, the Poisson's ratio is used to gauge how stable a crystal is under shear, and a high value of $\sigma$ indicates a high degree of plasticity [41]. For purely covalent materials, the value of $\sigma$ is around 0.10 and $G \sim 1.1B$, whereas, for ionic materials, the typical value of $\sigma$ is approximately 0.25 and $G \sim 0.6B$, and for metallic materials, $\sigma$ is typically 0.25 or above and $G \sim 0.4B$ [42]. These suggest that delocalized metallic bondings should be dominant in $Zr_3Ir$. Another method for determining the angular characteristics of bondings in a compound is the Cauchy pressure ($C''$). According to Pettifor's rule [43], a solid with high positive Cauchy pressures has more metallic bonds and is, therefore, more ductile, while a solid with large negative Cauchy pressures has more angular bonds and is, therefore, more brittle. Therefore, we can infer the ductility of $Zr_3Ir$ to the significant metallic bonding.

The term "machinability" refers to the properties of a material that indicate how easily it can be machined using a cutting tool. Information on a solid's machinability is essential in today's industry since it governs the optimum level of machine utilization, cutting forces, temperature, and plastic strain. This indicator can also be used as a measure of the plasticity and dry lubricating property of a solid [44–46]. A high value of $\mu_M$ suggests greater ease of shape manipulation, excellent dry lubricating properties, lower feed forces, lower friction value, and higher plastic strain value. The machinability index of $Zr_3Ir$ is found to be 3.63, indicating a very high level of machinability. At the same time the comound possesses significant dry lubricity.

The Kleinman parameter ($\zeta$), also known as the internal strain parameter, is an indicator that can be used to gauge how easily a substance bends and stretches.. The value of $\zeta$ typically ranges from 0 to 1 [25]. The lower and upper limits of $\zeta$ indicate the significant contribution of bond stretching/contracting to resist external loading and the significant contribution of bond bending to resist external loading, respectively. The calculated value of $\zeta$ of $Zr_3Ir$ is 0.99, indicating that the bond bending contribution primarily dominates $Zr_3Ir$'s mechanical strength.

**Table 3**

Calculated elastic moduli (all in GPa), Pugh's ratio ($B/G$), Poisson's ratio ($\sigma$), Cauchy pressure ($C''$) in GPa, machinability index ($\mu_M$), and Kleinman parameter ($\zeta$) of $Zr_3Ir$.

| Compound | $B$ | $G$ | $Y$ | $B/G$ | $\sigma$ | $C''$ | $\mu_M$ | $\zeta$ | Ref. |
|---|---|---|---|---|---|---|---|---|---|
| $Zr_3Ir$ | 129.28 | 26.78 | 75.15 | 4.83 | 0.403 | 101.45 | 3.63 | 0.99 | This work |
| | 128.80 | 28.60 | 79.90 | 4.50 | 0.397 | --- | --- | --- | [3]$^{\text{Theo.}}$ |

Table 3 includes the results from previous theoretical work [3]. Once again, very good agreement with the computed parameters are found.

Hardness is a property of solids that describes their resistance to localized plastic deformation. The hardness of the material is important from the application's point of view. Among the elastic constants and moduli, $C_{44}$ and $G$ are taken as the best hardness predictors of solids [47]. There are various proposed schemes to calculate the hardness theoretically. The hardness formulae developed by X. Chen *et al.* ($H_{\text{macro}}$) [48], Y. Tian *et al.* [$(H_{\text{v}})_{\text{Tian}}$] [49], and D. M. Teter [$(H_{\text{v}})_{\text{Teter}}$] [50] are based on either $G$ or both $G$ and $B$, whereas the formulae developed by N. Miao *et al.* ($H_{\text{micro}}$) [51] and E. Mazhnik *et al.* [$(H_{\text{v}})_{\text{Mazhnik}}$] [52] depend on the Young's modulus and Poisson's ratio. These formulae are as follows:

$$H_{micro} = \frac{(1-2\sigma)Y}{6(1+\sigma)} \qquad (19)$$

$$H_{macro} = 2[\left(\frac{G}{B}\right)^2 G]^{0.585} - 3 \quad (20)$$

$$(H_V)_{Tian} = 0.92(G/B)^{1.137}G^{0.708} \quad (21)$$

$$(H_V)_{Teter} = 0.151G \quad (22)$$

$$(H_V)_{\text{Mazhnik}} = \gamma_0\chi(\sigma)Y \quad (23)$$

In Equation (23), $\chi(\sigma)$ is a function of the Poisson's ratio and can be evaluated from:

$$\chi(\sigma) = \frac{1 - 8.5\sigma + 19.5\sigma^2}{1 - 7.5\sigma + 12.2\sigma^2 + 19.6\sigma^3}$$

where $\gamma_0$ is a dimensionless constant with a value of 0.096.

The difference in the hardness values obtained is due to the different parameters involved in the equations (19) – (23). These values are disclosed in Table 4. All the computed values are low, and the value for $H_{\text{macro}}$ proposed by X. Chen *et al.* [48] is found to be negative, indicating that $Zr_3Ir$ is significantly soft material.

**Table 4**

Calculated hardness (GPa) based on elastic moduli and Poisson's ratio of $Zr_3Ir$.

| Compound | $(H_V)_{Micro}$ | $(H_V)_{macro}$ | $(H_V)_{Tian}$ | $(H_V)_{Teter}$ | $(H_V)_{Mazhnik}$ |
|---|---|---|---|---|---|
| $Zr_3Ir$ | 1.73 | -0.83 | 1.58 | 4.04 | 4.32 |

The elastic anisotropy of crystal explains the directional dependence of the mechanical properties of a compound and is an important parameter for engineering science and for estimating the mechanical properties of a compound under different conditions of external loadings. Moreover, it is important to calculate the elastic anisotropy in a material in order to understand a number of physical properties such as the formation and propagation of microscale cracks in ceramics, plastic relaxation in thin films, etc. The following widely used formulae are used to calculate anisotropy indicators of $Zr_3Ir$:

The shear anisotropy factor for a tetragonal crystal can be quantified by three different parameters [53,54]: Considering the {100} shear planes between the ⟨011⟩ and ⟨010⟩ directions, the shear anisotropy factor, $A_1$ is,

$$A_1 = \frac{4C_{44}}{C_{11}+C_{33}-2C_{13}} \quad (24)$$

Considering the {010} shear plane between ⟨101⟩ and ⟨001⟩ directions the shear anisotropy factor, $A_2$ is,

$$A_2 = \frac{4C_{55}}{C_{22} + C_{33} - 2C_{23}} \qquad (25)$$

Considering the {001} shear planes between ⟨110⟩ and ⟨010⟩ directions, the anisotropy factor, $A_3$ is,

$$A_3 = \frac{4C_{66}}{C_{11} + C_{22} - 2C_{12}} \qquad (26)$$

In the case of an isotropic crystal, the factors $A_1$, $A_2$, and $A_3$ must be equal to one, while any value smaller or greater than one is a measure of the degree of elastic anisotropy possessed by the crystal for shape changing deformation. The calculated values of these anisotropy factors are given in Table 5. The estimated values of $A_1$, $A_2$, and $A_3$ predict that the compound $Zr_3Ir$ is moderately anisotropic. $A_1$and $A_2$ are equal as the compound $Zr_3Ir$ belongs to the tetragonal system.

The universal log-Euclidean anisotropy index is defined as follows [55,56]:

$$A^L = \sqrt{[\ln(\frac{B_V}{B_R})]^2 + 5\,[\ln(\frac{C_{44}^V}{C_{44}^R})]^2} \qquad (27)$$

In this scheme, the Voigt and Reuss approximated values of $C_{44}$ is calculated from [57]:

$$C_{44}^V = C_{44}^R + \frac{3}{5}\frac{(C_{11} - C_{12} - 2C_{44})^2}{3(C_{11} - C_{12}) + 4C_{44}}$$

and

$$C_{44}^R = \frac{5}{3}\frac{C_{44}(C_{11} - C_{12})}{3(C_{11} - C_{12}) + 4C_{44}}$$

For perfect isotropy, $A^L = 0$. Considering only the value of $A^L$, it is difficult to ascertain whether a solid is layered/lamellar or not. Generally, majority (78%) of inorganic crystalline solids with high $A^L$ values, exhibit layered/lamellar structure [57]. In general, compounds having higher (lower) $A^L$ values show strong layered (non-layered) structural features. For the high value of $A^L$, $Zr_3Ir$ is expected to exhibit layered feature.

The universal anisotropy index, $(A^U, d_E)$, equivalent Zener anisotropy measure, $A^{eq}$, percentage anisotropy in compressibility, $A_B$ and anisotropy in shear, $A_G$ (or $A_C$) for crystals are calculated from the following expressions [55,58,59]:

$$A^U = 5\frac{G_V}{G_R} + \frac{B_V}{B_R} - 6 \geq 0 \qquad (28)$$

$$d_E = \sqrt{A^U + 6} \qquad (29)$$

$$A^{eq} = \left(1 + \frac{5}{12}A^U\right) + \sqrt{(1 + \frac{5}{12}A^U)^2 - 1} \quad (30)$$

$$A_B = \frac{B_V - B_R}{B_V + B_R} \quad (31)$$

$$A_G = \frac{G_V - G_R}{2G_H} \quad (32)$$

The universal anisotropy index [58], $A^U$, is a singular measure of anisotropy and applicable for all crystal systems irrespective of their symmetry. $A^U = 0$ implies isotropy, while nonzero value of $A^U$ indicates anisotropy. The estimated value of $A^U$ predicts that $Zr_3Ir$ possesses anisotropy. On the other hand, equivalent Zener anisotropy measure, $A^{eq}= 1$, represents isotropy, while any other value suggests anisotropy. The calculated values of $A^{eq}$ for $Zr_3Ir$ is given in Table 5, also predicting that $Zr_3Ir$ is elastically anisotropic. $A_B = A_G = 0$ represent perfect elastic isotropy, while $A_B = A_G = 1$, represents maximum elastic anisotropy. Compared to $A_B$, the larger value of $A_G$ (Table 5) indicates that anisotropy in shear is greater than the anisotropy in compressibility for $Zr_3Ir$.

The linear compressibility of a tetragonal crystal along *a*- and *c*-axis ($\beta_a$ and $\beta_c$) are calculated using [60,61]:

$$\beta_a = \frac{C_{33} - C_{13}}{D} \quad (33)$$

$$\beta_c = \frac{C_{11} + C_{12} - 2C_{13}}{D} \quad (34)$$

with $D = (C_{11} + C_{12})C_{33} - 2(C_{13})^2$

The calculated values (see Table 5) for $Zr_3Ir$ indicate that compressibility along *a*-axis is lower than that along *c*-axis, in agreement with the elastic constants.

**Table 5**

Shear anisotropy factors ($A_1$, $A_2$ and $A_3$), universal log-Euclidean index $A^L$, the universal anisotropy index $(A^U, d_E)$, equivalent Zener anisotropy measure $A^{eq}$, anisotropy in shear $A_G$, anisotropy in compressibility $A_B$, linear compressibilities ($\beta_a$ and $\beta_c$) (TPa$^{-1}$) and their ratio, $\frac{\beta_c}{\beta_a}$, for $Zr_3Ir$.

| Compound | $A_1$ | $A_2$ | $A_3$ | $A^L$ | $A^U$ | $d_E$ | $A^{eq}$ | $A_G$ | $A_B$ $(10^{-3})$ | $\beta_a$ $(10^{-3})$ | $\beta_c$ $(10^{-3})$ | $\frac{\beta_c}{\beta_a}$ |
|---|---|---|---|---|---|---|---|---|---|---|---|---|
| $Zr_3Ir$ | 1.00 | 1.00 | 1.80 | 1.48 | 1.20 | 2.68 | 2.62 | 0.11 | 1.70 | 2.36 | 3.03 | 1.28 |

The directional bulk moduli along $a$-, $b$- and $c$-axis in the crystal (giving the bulk modulus under uniaxial strain) and anisotropies of the bulk modulus are calculated from [62]:

$$B_a = a\frac{dP}{da} = \frac{\Lambda}{1 + \alpha + \beta} \quad (35)$$

$$B_b = a\frac{dP}{db} = \frac{B_a}{\alpha} \quad (36)$$

$$B_c = c\frac{dP}{dc} = \frac{B_a}{\beta} \quad (37)$$

$$A_{B_a} = \frac{B_a}{B_b} = \alpha \quad (38)$$

$$A_{B_c} = \frac{B_c}{B_b} = \frac{\alpha}{\beta} \quad (39)$$

where, $\Lambda = C_{11} + 2C_{12}\alpha + C_{22}\alpha^2 + 2C_{13}\beta + C_{33}\beta^2 + 2C_{33}\alpha\beta$, and for a tetragonal solid, $\alpha$ = 1 and $\beta = \frac{C_{11}+C_{12}-2C_{13}}{C_{33}-C_{13}}$. $A_{B_a}$ and $A_{B_c}$ are the anisotropies of bulk moduli along $a$-axis and $c$-axis with respect to the $b$-axis, respectively. The calculated values are presented in Table 6. $A_{B_a} = 1$ and $A_{B_c} \neq 1$, indicate that the directional bulk modulus for $Zr_3Ir$ is anisotropic. From Table 6, it is also seen that bulk modulus along $c$-axis is smaller than those along $a$- and $b$-axis. All these values of constrained bulk moduli are higher than that of the isotropic polycrystalline bulk modulus.

**Table 6**

Anisotropies of bulk modulus along different crystallographic axes of $Zr_3Ir$.

| Compound | $B_a$ | $B_b$ | $B_c$ | $A_{B_a}$ | $A_{B_c}$ |
|---|---|---|---|---|---|
| $Zr_3Ir$ | 486.82 | 486.82 | 377.38 | 1 | 0.78 |

We have analyzed and shown the ELATE [63] generated two-dimensional (2D) and three-dimensional (3D) profiles of the Young modulus, shear modulus, linear compressibility (inverse of the bulk modulus), and Poisson's ratio of the binary intermetallic $Zr_3Ir$ compound. The 3D contour plots must have spherical shapes for mechanically isotropic solids; otherwise, anisotropy is present. The 3D profiles of *Y*, *β*, *G*, and *σ* are shown in Figure 2, all of which deviate somewhat from the spherical shape, indicating some degree of anisotropy. ELATE additionally provides a numerical investigation by presenting the minimum and maximum values of each modulus as well as the directions along which these extrema occur. The maximum and minimum values of *Y*, *β*, *G*, and *σ* and their maximum to minimum ratios are disclosed in Table 7. These ratios are useful factors to define the elastic anisotropy in $Zr_3Ir$. The projections of the 3D profiles on the principal crystal planes are shown in the 2D plots. These plots illustrates the anisotropy in the elastic properties in different directions within the specific crystal planes. It is seen that the compressibility (and thus, the bulk modulus, is completely isotropic in the *xy*(*ab*)-plane. This is a consequence of the tetragonal symmetry of $Zr_3Ir$.

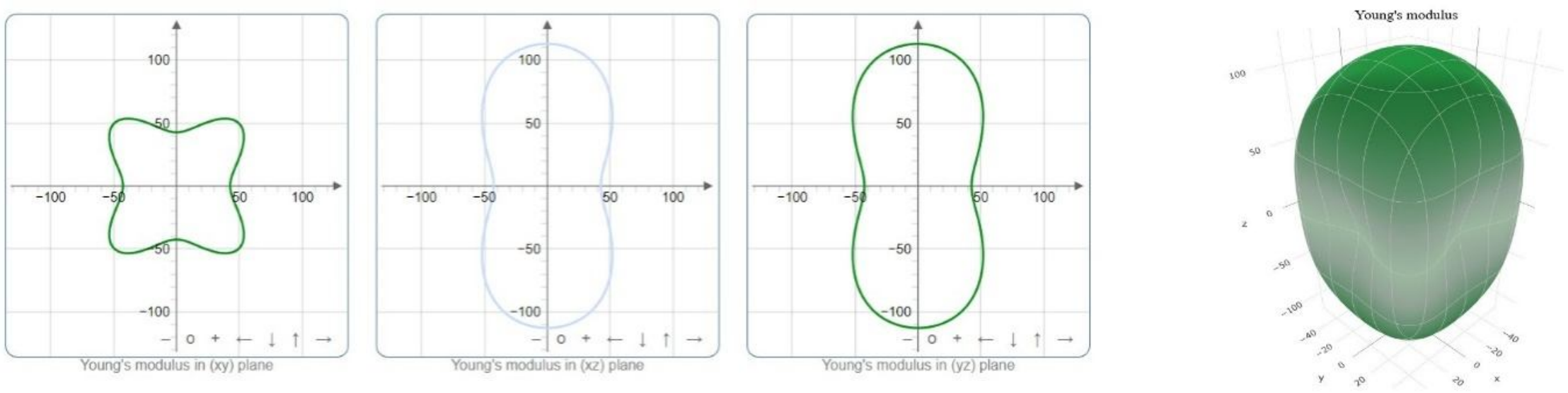

(a)

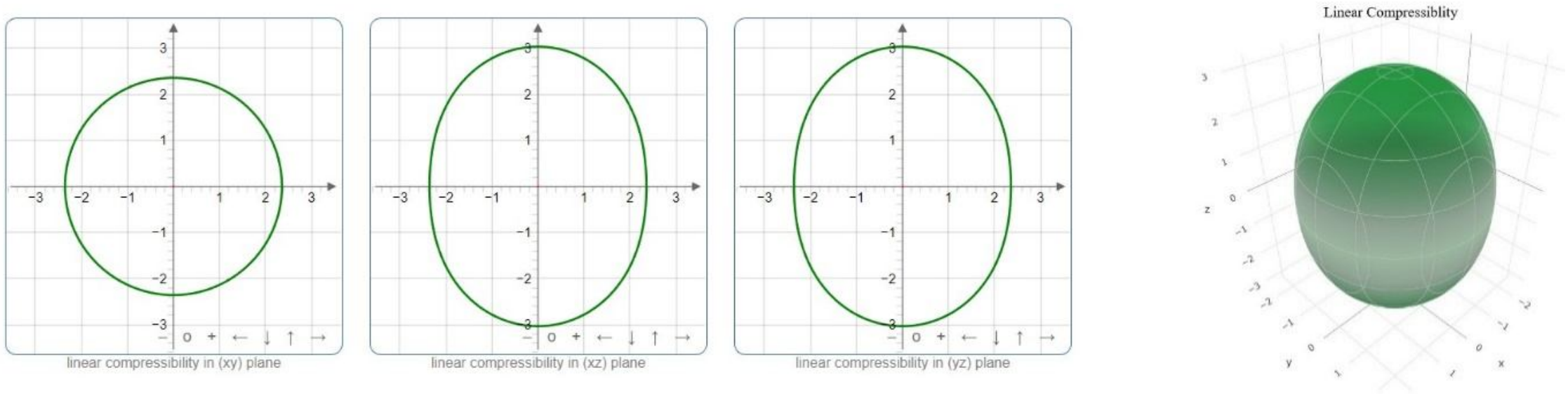

(b)

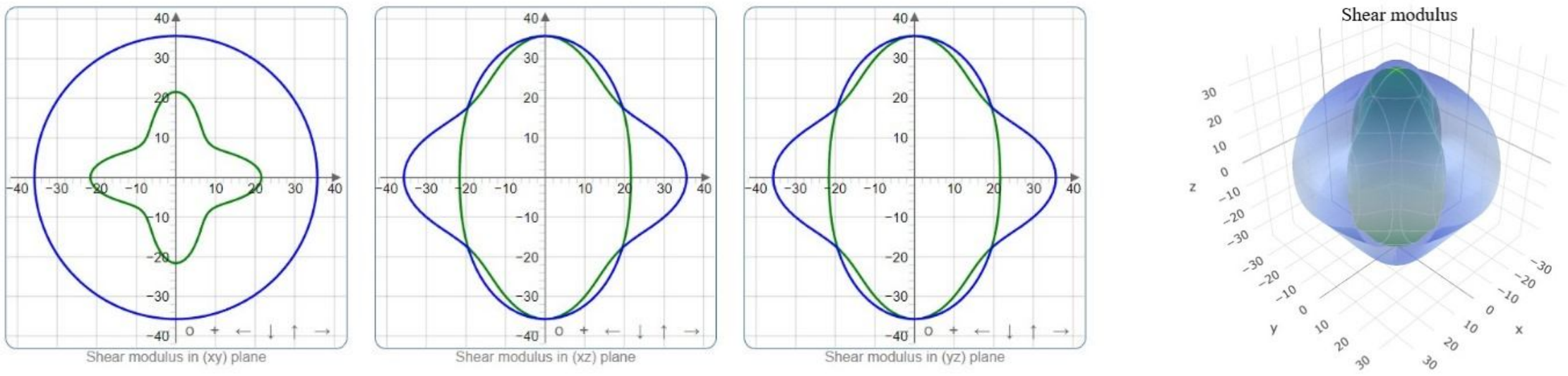

(c)

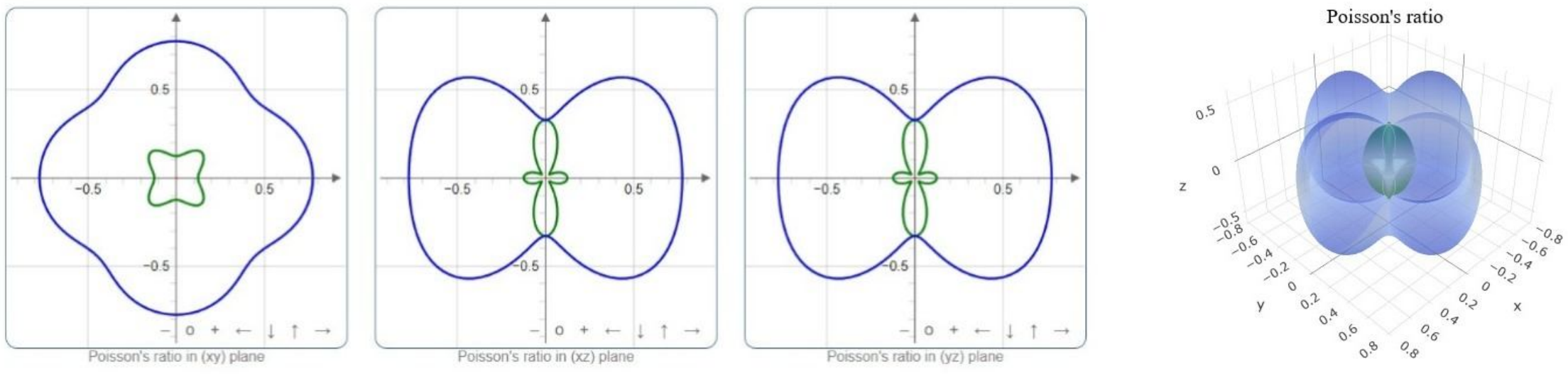

(d)

**Figure 2.** 2D (left) and 3D (right) directional dependences of (a) Young modulus (b) linear compressibility (c) shear modulus and (d) Poisson's ratio for $Zr_3Ir$.

From Table 7, it is observed that the Poisson's ratio is highly anisotrpic. This is a consequence of the highly layered structure of $Zr_3Ir$.

**Table 7**

The minimum and maximum values of Young's modulus (GPa), compressibility ($Tpa^{-1}$), shear modulus (Gpa), Poisson's ratio, and their ratios for $Zr_3Ir$.

| Phase | $Y$ | | $A_Y$ | $\beta$ | | $A_\beta$ | $G$ | | $A_G$ | $\sigma$ | | $A_\sigma$ |
|---|---|---|---|---|---|---|---|---|---|---|---|---|
| | $Y_{min}$ | $Y_{max}$ | | $\beta_{min}$ | $\beta_{max}$ | | $G_{min}$ | $G_{max}$ | | $\sigma_{min}$ | $\sigma_{max}$ | |
| $Zr_3Ir$ | 42.59 | 112.88 | 2.65 | 2.36 | 3.03 | 1.29 | 11.99 | 35.61 | 2.97 | 0.03 | 0.80 | 28.18 |

### *3.3. Charge density distribution*

To comprehend the atomic bondings between the atoms of $Zr_3Ir$ directly, the valence electron charge density distributions within the (100) and (010) planes are displayed in Figure 3. The color scale at the right of the maps shows the total electron density. The charge density distribution map indicates the ionic bonds between Zr-Zr, Zr-Ir, and Ir-Ir atoms. The charge density contours are almost circular for all the atoms in different planes. This is a strong evidence that covalent bondings are very weak in this compound. According to the color scale, red denotes a high electron charge density whereas blue denotes a low electron charge density. So, Zr atoms have a high electron density compared to Ir atoms. As can be observed from the charge density distribution maps, there is weak localization of charge surrounding the Zr and Ir atoms. This is suggestive of weak covalent bonding, a result consistent with the bond population analysis (section 3.4.). The uniform blue region with smeared electronic charge is consistent with the metallic bonding.

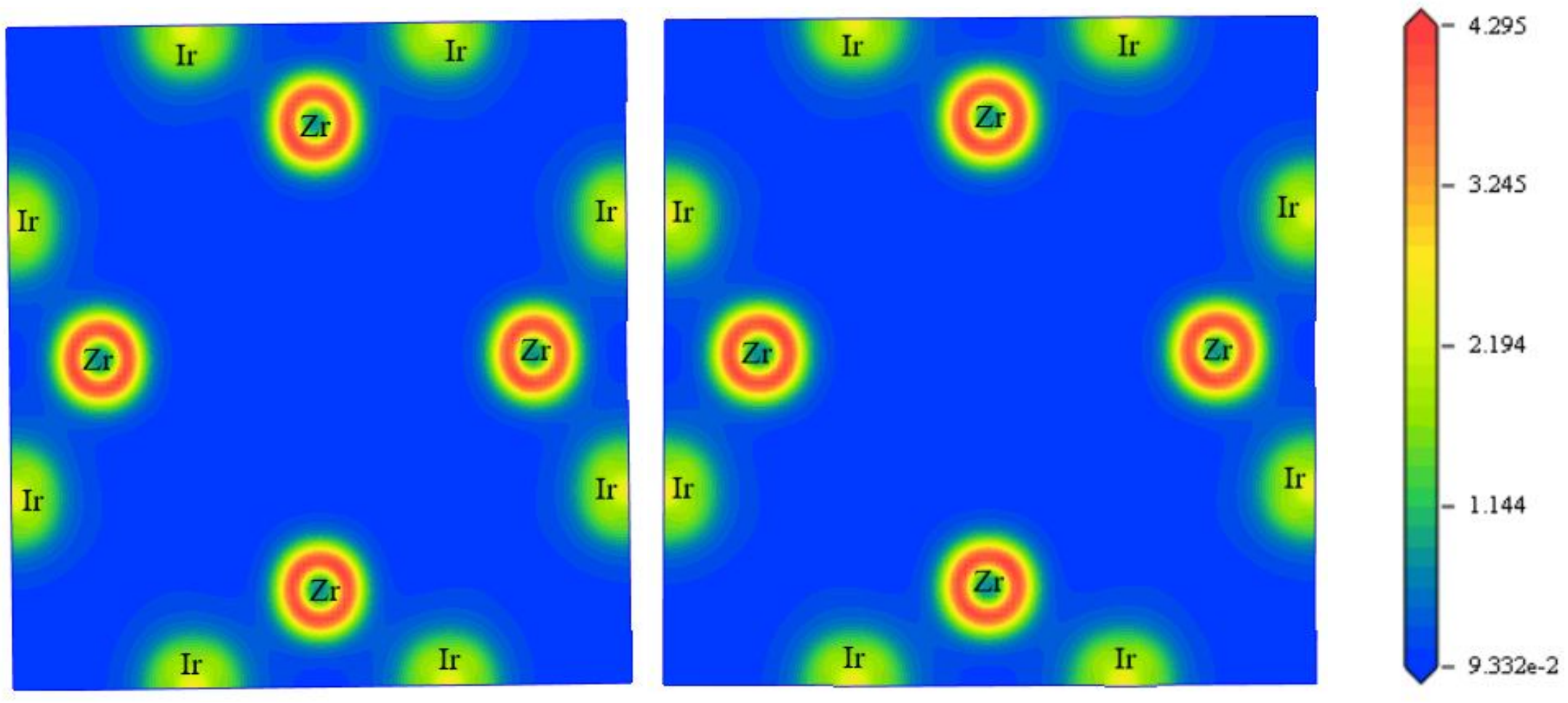

(a) (b)

**Figure 3.** Charge density distribution map of $Zr_3Ir$ in (a) (100) and (b) (010) planes. The charge density scale is given at the right of the panels.

### *3.4. Bond population analysis*

Both Mulliken population analysis (MPA) [17] and Hirshfeld population analysis (HPA) [64] has been performed to investigate the bonding type (ionic, covalent, or metallic) and the effective valence of an atom in $Zr_3Ir$ compound. The results of these analyses are listed in Table 8. The charge spilling parameter, which specifies how much of valence charge is absent from a projection, is also tabulated in Table 8. The lower value of the charge spilling parameter indicates an efficient representation of the electronic bonds. According to Table 8, the total charge for Zr atoms is significantly higher than that for Ir atoms and which mainly comes from the 4*p* states of Zr. The difference between the formal ionic charge and the calculated Mulliken charge for an element is referred to as the effective valence charge (EVC) [19]. A non-zero value of EVC indicates the level of covalency. It is known that MPA often overestimates the covalent contribution in the overall bonding. To get more meaningful results, we have also calculated HPA. The results from HPA are qualitatively consistent with those of the MPA but the level of covalency is much lower than those of the MPA.

**Table 8**

Charge spilling parameter (%), orbital charges (electron), atomic Mulliken charge (electron), Hirshfeld charge (electron), and EVC (electron) of $Zr_3Ir$.

| Compound | Charge spilling (%) | Atoms | s | p | d | Total | Formal ionic charge | Mulliken charge | Hirshfeld charge | EVC |
|---|---|---|---|---|---|---|---|---|---|---|
| $Zr_3Ir$ | 0.09 | Zr | 2.35 | 6.69 | 2.80 | 11.84 | +4 | 0.16 | 0.04 | 3.84 |
| | | | 2.33 | 6.72 | 2.78 | 11.83 | +4 | 0.17 | 0.06 | 3.83 |
| | | | 2.33 | 6.66 | 2.71 | 11.69 | +4 | 0.31 | 0.11 | 3.69 |
| | | Ir | 0.80 | 1.03 | 7.81 | 9.64 | +3 | -0.64 | -0.21 | 3.64 |

### *3.5. Theoretical bond hardness and fracture toughness*

The hardness of a solid plays an important role in its applications, particularly as an abrasive resistant phase and radiation tolerant material [65]. A solid is harder if it has a higher bond population or electronic density and shorter bond length. F. Gao [66] and H. Gou *et al.* [67]. established the Vickers hardness, which is a well-known theoretical method for determining intrinsic hardness. The computed Vickers hardness is mostly dependent on the Mulliken overlap population, which gives the degree of overlap of the electron clouds forming bonds between atoms within the crystal. The overlap population of electrons between atoms is a measure of the strength of the bond per unit volume. The hardness of a metallic material should take account of the metallic bonding and can be obtained using the following equations [66,68]:

$$H_v^\mu = \left[\prod^{\mu}\left\{740(P^\mu - P^{\mu\prime})\left(v_b^\mu\right)^{-5/3}\right\}^{n^\mu}\right]^{1/\sum n^\mu} \quad (40)$$

and

$$H_v = \left[\prod^{\mu}(H_v^\mu)^{n^\mu}\right]^{1/\sum n^\mu} \quad (41)$$

In these equations, $P^\mu$ is the Mulliken bond overlap population of the $\mu$-type bond, $P^{\mu'} = n_{free}/V$ is the metallic population density, where $n_{free} = \int_{E_P}^{E_F} N\,(E)\,dE$ = the number of free electrons; $E_P$ and $E_F$ are the energy at the pseudogap and at the Fermi level, respectively, $n^\mu$ is the number of $\mu$-type bond, $v_b^\mu$ is the bond volume of $\mu$-type bond calculated from the equation $v_b^\mu = (d^\mu)^3/\sum_v[(d^\mu)^3 N_b^\mu]$, $H_v^\mu$ is the bond hardness of the $\mu$-type bond and $H_v$ is the bulk hardness of the solid. The constant 740 is a proportionality coefficient extracted by fitting the hardness of diamond [66,68].

The calculated bond length, overlap population, and the theoretical hardness of $Zr_3Ir$ are presented in Table 9. The value of Vickers hardness for $Zr_3Ir$ is 4.94 GPa.

Mulliken population analysis is also helpful in determining the metallic nature of the bonds in a compound. Metallic bonding is relatively softer compared to ionic and covalent ones [69]. The metallicity of a crystal can be found from:

$$f_m = {P^{\mu'}}/{P^\mu} \quad (42)$$

where, $P^{\mu'}$ and $P^\mu$ are the metallic population and the Mulliken bond overlap population, respectively. The calculated values are disclosed in Table 9. The calculated values of hardness $H_v^\mu$ of a μ-type bond and total hardness $H_v$ of $Zr_3Ir$ are all summarized in Table 9.

One of the major problems with the surface hard coatings on large machinery is the development of cracks, especially in metals and ceramic materials. Fracture toughness, $K_{IC}$ is a parameter, which can evaluate the resistance of a material against this crack/fracture propagation. The formula for $K_{IC}$ of a material is as follows [70]:

$$K_{IC} = \alpha_0^{-1/2} V_0^{1/6}[\xi(\sigma)Y]^{3/2} \quad (43)$$

where $V_0$ = volume per atom; $\alpha_0$ = 8840 GPa for covalent and ionic crystals; $\xi(\sigma)$ is a dimensionless parameter function of the Poisson's ratio $(\sigma)$, which can be found from:

$$\xi(\sigma) = \frac{1 - 13.7\sigma + 48.6\sigma^2}{1 - 15.2\sigma + 70.2\sigma^2 - 81.5\sigma^3}$$

As presented in Table 9, the value of $K_{IC}$ of $Zr_3Ir$ is 1.71 MPam$^{1/2}$. The ratio of hardness and fracture toughness is called an index of brittleness ($B_i$) of a solid which expresses the degree of damage tolerance. The lower value of $B_i$ indicates higher damage tolerance of materials [71]. The value of $B_i$ for $Zr_3Ir$ is 2.89 which predicts a high damage tolerance of $Zr_3Ir$.

**Table 9.**

The calculated Mulliken bond overlap population of $\mu$-type bond $P^{\mu}$, bond length $d^{\mu}$ (Å), metallic population $P^{\mu'}$, metallicity $f_m$, the total number of bonds $N^{\mu}$, cell volume $\Omega$ (Å$^3$), bond volume $v_b^{\mu}$ (Å$^3$), hardness of the $\mu$-type bond $H_v^{\mu}$ (GPa), Vickers hardness of the compound $H_v$ (GPa), fracture toughness, $K_{IC}$ (MPam$^{1/2}$) and index of brittleness of $Zr_3Ir$.

| Compound | Bond | | $d^{\mu}$ | $P^{\mu}$ | $P^{\mu'}$ | $f_m$ | $N^{\mu}$ | Ω | $v_b^{\mu}$ | $H_v^{\mu}$ | $H_v$ | $K_{IC}$ | $H_v/K_{IC}$ |
|---|---|---|---|---|---|---|---|---|---|---|---|---|---|
| | Zr24-Ir3 | | | | | | | | | | | | |
| | Zr23-Ir4 | | | | | | | | | | | | |
| | Zr21-Ir4 | | | | | | | | | | | | |
| | Zr19-Ir6 | | | | | | | | | | | | |
| | Zr18-Ir5 | | | | | | | | | | | | |
| | Zr17-Ir6 | | | | | | | | | | | | |
| | Zr24-Ir1 | | | | | | | | | | | | |
| | Zr18-Ir7 | Zr-Ir | 2.74 | 0.45 | 0.01 | 0.02 | 16 | 664.703 | 10.31 | 6.67 | | | |
| | Zr22-Ir3 | | | | | | | | | | | | |
| | Zr20-Ir5 | | | | | | | | | | | | |
| | Zr22-Ir1 | | | | | | | | | | | | |
| | Zr20-Ir7 | | | | | | | | | | | | |
| | Zr23-Ir2 | | | | | | | | | | | | |
| | Zr21-Ir2 | | | | | | | | | | | | |
| | Zr19-Ir8 | | | | | | | | | | | | |
| | Zr17-Ir8 | | | | | | | | | | | | |
| | Zr15-Ir8 | | | | | | | | | | | | |
| | Zr11-Ir2 | | 2.79 | 0.34 | 0.01 | 0.03 | 16 | 664.703 | 10.88 | 4.57 | | | |
| | Zr13-Ir7 | | | | | | | | | | | | |
| | Zr13-Ir3 | Zr-Ir | | | | | | | | | | | |
| | Zr9-Ir5 | | | | | | | | | | | | |
| | Zr9-Ir1 | | | | | | | | | | | | |
| | Zr5-Ir5 | | | | | | | | | | | | |
| | Zr5-Ir3 | | | | | | | | | | | | |
| | Zr3-Ir8 | | | | | | | | | | | | |
| | Zr3-Ir2 | | | | | | | | | | | | |
| | Zr1-Ir7 | | | | | | | | | | | | |
| | Zr1-Ir1 | | | | | | | | | | 4.94 | 1.71 | 2.89 |
| | Zr15-Ir4 | | | | | | | | | | | | |
| | Zr11-Ir6 | | | | | | | | | | | | |
| $Zr_3Ir$ | Zr7-Ir6 | | | | | | | | | | | | |
| | Zr7-Ir4 | | | | | | | | | | | | |
| | Zr8-Ir7 | | | | | | | | | | | | |
| | Zr6-Ir8 | | | | | | | | | | | | |
| | Zr6-Ir2 | | | | | | | | | | | | |
| | Zr14-Ir2 | | | | | | | | | | | | |
| | Zr10-Ir8 | | | | | | | | | | | | |
| | Zr8-Ir1 | | | | | | | | | | | | |
| | Zr2-Ir6 | | | | | | | | | | | | |
| | Zr2-Ir4 | | | | | | | | | | | | |
| | Zr16-Ir5 | | | | | | | | | | | | |
| | Zr12-Ir3 | | | | | | | | | | | | |
| | Zr14-Ir6 | Zr-Ir | 2.79 | 0.30 | 0.01 | 0.03 | 16 | 664.703 | 10.90 | 4.005 | | | |
| | Zr10-Ir4 | | | | | | | | | | | | |
| | Zr4-Ir5 | | | | | | | | | | | | |
| | Zr4-Ir3 | | | | | | | | | | | | |
| | Zr16-Ir1 | | | | | | | | | | | | |
| | Zr12-Ir7 | | | | | | | | | | | | |
| | Zr11-Zr15 | Zr-Zr | 2.92 | 0.25 | 0.01 | 0.04 | 4 | 664.703 | 12.53 | 2.63 | | | |
| | Zr3-Zr7 | | | | | | | | | | | | |
| | Zr1-Zr5 | | | | | | | | | | | | |
| | Zr9-Zr13 | | | | | | | | | | | | |
| | Zr22-Ir6 | Zr-Ir | 2.93 | 0.62 | 0.01 | 0.02 | 8 | 664.703 | 12.64 | 6.58 | | | |
| | Zr24-Ir8 | | | | | | | | | | | | |
| | Zr23-Ir7 | | | | | | | | | | | | |
| | Zr17-Ir1 | | | | | | | | | | | | |
| | Zr21-Ir5 | | | | | | | | | | | | |
| | Zr20-Ir4 | | | | | | | | | | | | |
| | Zr19-Ir3 | | | | | | | | | | | | |
| | Zr18-Ir2 | | | | | | | | | | | | |

From Table 9, it is seen that Zr-Ir bondings have different levels of bond hardness depending on the location of the atoms within the crystal. The lowest bond hardness value is found for the Zr-Zr bondings which has the largest level of metallicity.

### *3.6. Electronic band structure and density of states*

The valence and conduction electrons present in solids control almost all of their physically structural, electronic, and optical properties. Certain magnetic orders and Pauli paramagnetic susceptibility are also controlled by the band structure and electronic energy density of states at the Fermi level of a material. How these electrons behave within the Brillouin zone depends on the characteristics of their energy dispersion ($E(k)$). This variation of momentum-based electronic energy in various bands provides as a representation of the electronic band structure. We have calculated the electronic band structure for the optimized crystal structure of $Zr_3Ir$ along several high symmetry directions (*Z-A-M-Γ-Z-R-X-Γ*) in the first BZ. The band structure is depicted in Figure 4. The horizontal broken line indicates the Fermi level. The charateristics of band structure clearly demonstrate that a number of bands with varying degrees of dispersion cross the Fermi level, as indicated by the green colored dispersion lines. This confirms the metallic character of $Zr_3Ir$. Highly dispersive bands results in a low charge carrier effective mass [72–74] and high charge mobility. For $Zr_3Ir$, the bands along *M-Γ*, *Γ-Z*, and *X-Γ* directions are highly dispersive, indicating lower effective mass of the charge carriers, consequently, a high mobility for the same. The curves along *A-M* direction are less dispersive, indicating a relatively higher effective mass of charge carriers, which results in a reduced mobility. It is worth noting that the bands crossing near the *Γ*-point exhibit a hole-like character for $Zr_3Ir$.

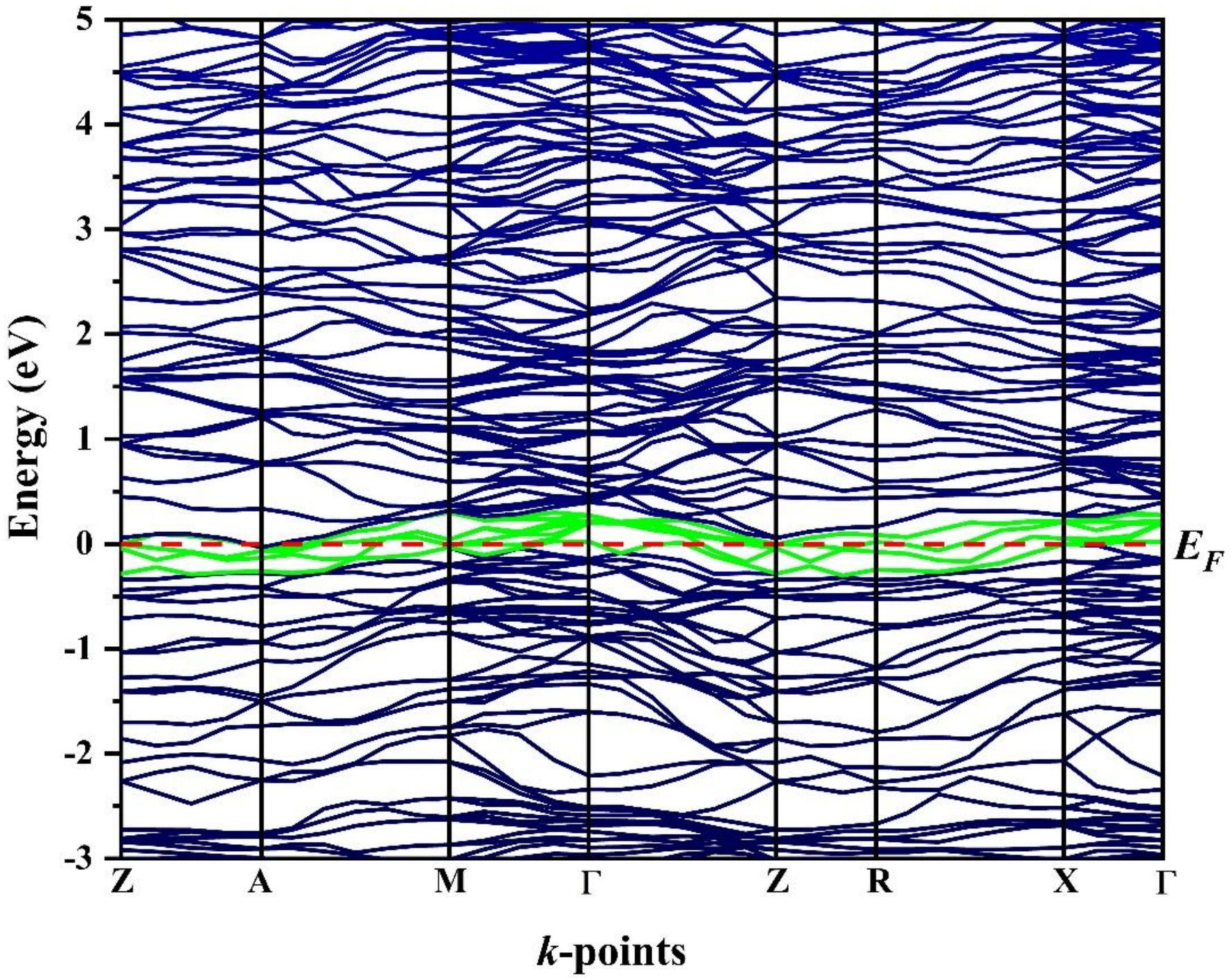

**Figure 4.** The electronic band structure of $Zr_3Ir$ along the high symmetry directions within the first Brillouin zone.

We have calculated the total density of states (TDOS) and partial density of states (PDOS) of $Zr_3Ir$ to better highlight the contributions of various orbitals to electronic characteristics and the nature of chemical bonding, as shown in Figure 5. The vertical dashed line at 0 eV represents the Fermi level, $E_F$. A finite value of TDOS (4.83 states/eV/formula unit) at the Fermi level suggests the metallic electrical conductivity of $Zr_3Ir$. To investigate how each atom contributed to the TDOS, we also computed the PDOS of $Zr_3Ir$. The Zr-4*d* states makes a major contribution (67.37%) to the TDOS at the $E_F$, which should dominate the electrical conductivity of $Zr_3Ir$. The properties of this electronic state also have an impact on the chemical and mechanical stabilities of $Zr_3Ir$. There is a significant overlap among the Zr-4*d*, Zr-4*p*, Ir-5*p*, and Ir-5*d* states in $Zr_3Ir$. This overlapping of the electronic states result in the formation of atomic bondings. In the TDOS, there are several peaks, e.g., at ~ - 3.0 eV, - 0.5 eV, 2 eV, 5 eV, 6.5 eV, and 10.0 eV. All of these peaks are expected to contribute significantly in optical transitions. Especially, the peaks near the Fermi level should regulate the movement of charges and related electrical transport properties. The nearest peak at the negative energy below the Fermi level in the TDOS is known as the bonding peak, while the nearest peak at the positive energy is the anti-bonding peak. The energy gap between these peaks is called the pseudo-gap which is an indication of electrical stability [75–77]. Bonding and anti-bonding peaks for $Zr_3Ir$ are located within 2.5 eV from the Fermi level. The Fermi level is located to the left of the pseudogap indicating that the bonding orbitals of $Zr_3Ir$ are partially filled. This also suggests that application of pressure and/or suitable atomic substitution can change the electronic and structural properties of $Zr_3Ir$ significantly.

The electron-electron interaction parameter of a material, termed as the repulsive Coulomb pseudopotential, can be calculated as follows from the electronic density of states at the Fermi energy [78]:

$$\mu^* = \frac{0.26\, N(E_F)}{1 + N(E_F)} \quad (44)$$

where, $N(E_F)$ is the total density of states at the Fermi level of the compound. The electron-electron interaction parameter of $Zr_3Ir$ is therefore found to be 0.22. The repulsive Coulomb pseudopotential is responsible to lower the transition temperature, $T_c$, of superconducting compounds [78,79]. It also measures the degree of electron correlations in a system. Relatively high value of this parameter is a consequence of the high TDOS at the Fermi level mainly due to the 4*d* electrons of Zr atoms.

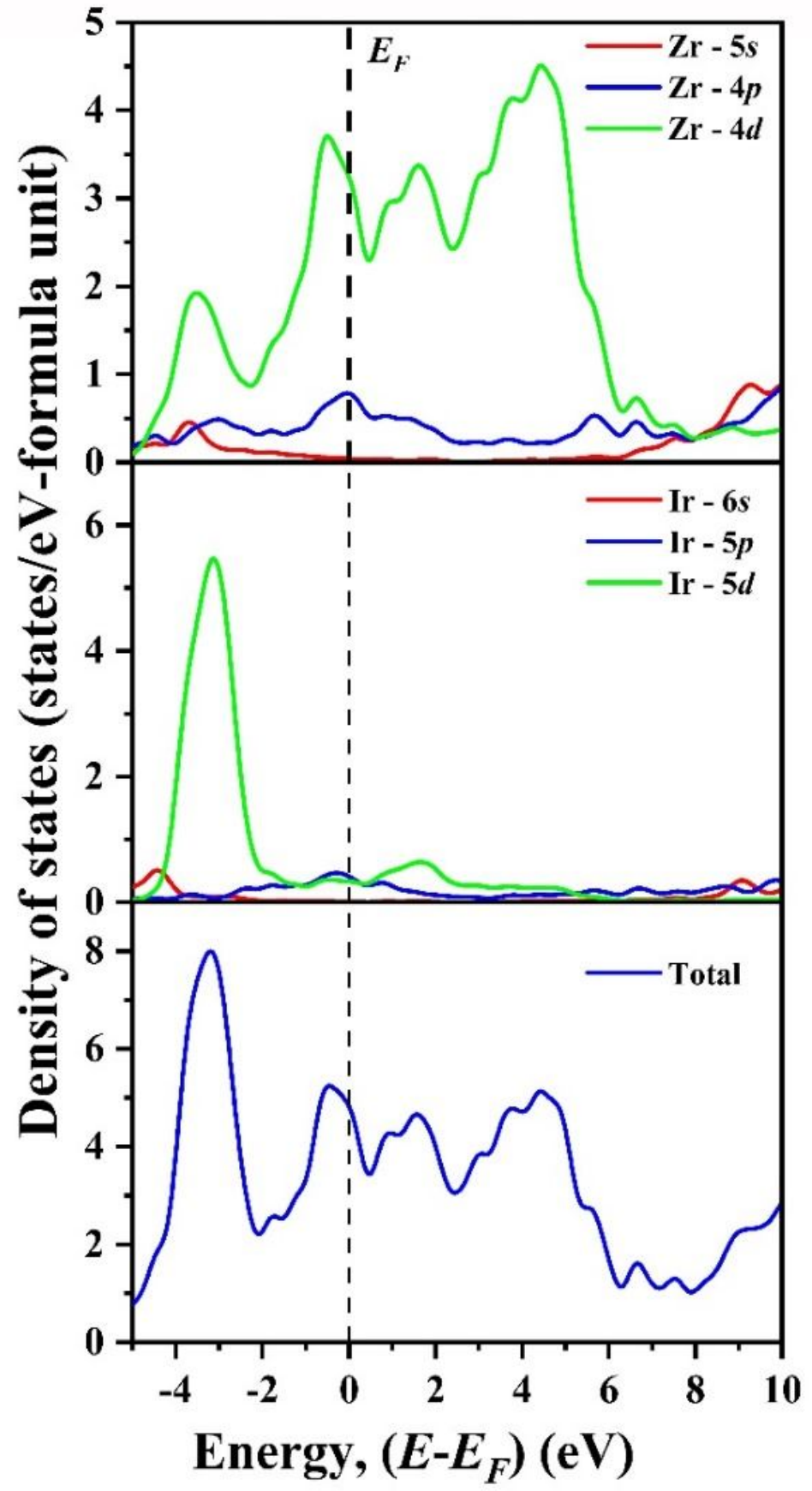

**Figure 5.** Total and partial electronic energy density of states of $Zr_3Ir$.

### *3.7. Fermi surface of $Zr_3Ir$*

In solid state physics, the study of the Fermi surface of a metallic system is important to understand the behavior of occupied and unoccupied electronic states at low temperatures.

Many electrical, transport, optical, and magnetic properties of a material are heavily influenced by its Fermi surface topology. As illustrated in Figure 6, the Fermi surfaces of $Zr_3Ir$ have been constructed from the corresponding electronic band structures. Fermi surfaces are constructed form with band numbers 178, 179, 180, 181, 182, 183, and 184, which cross the Fermi level.

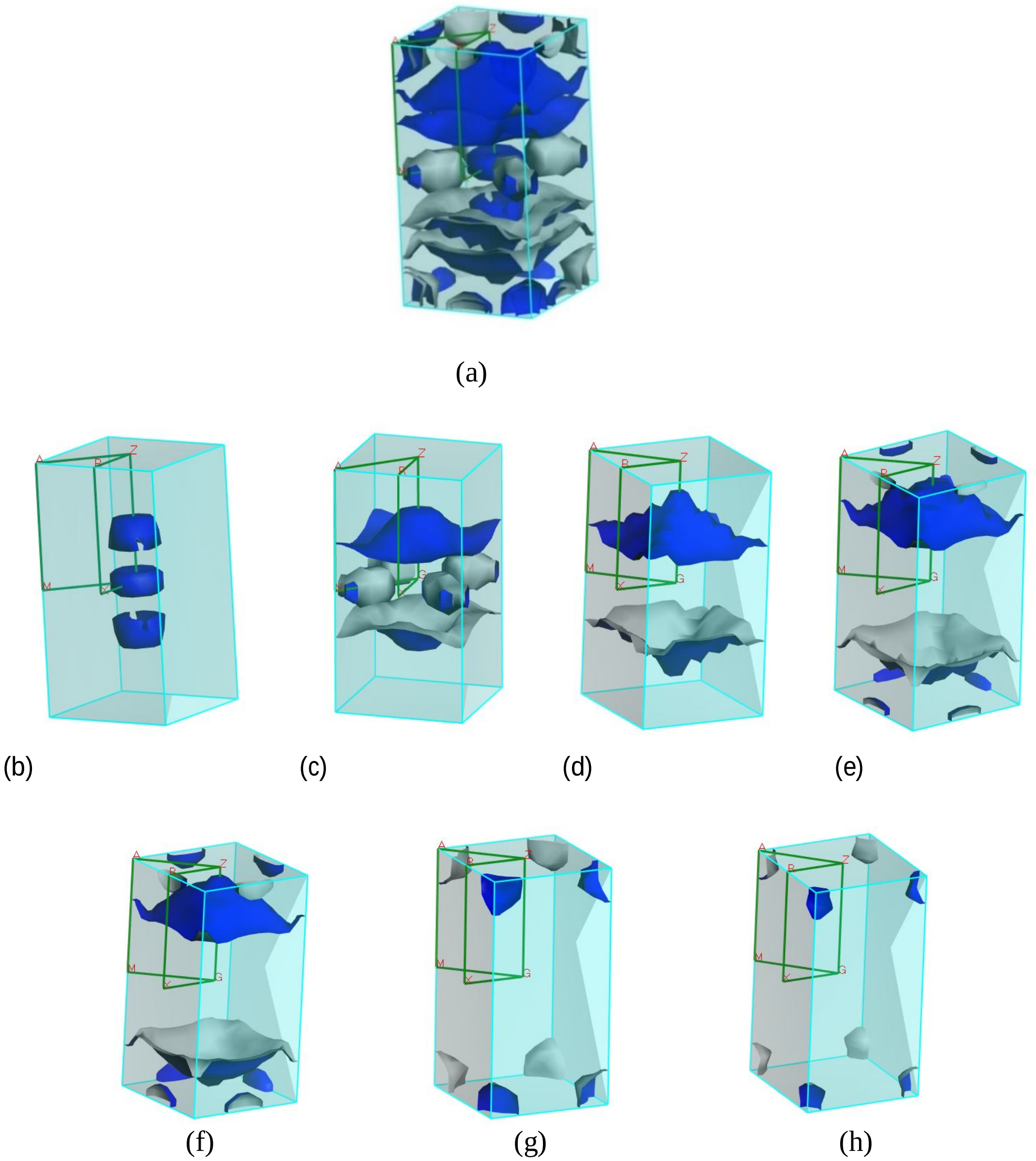

**Figure 6.** Fermi surface diagram for bands (a) all (b) 178 (c) 179 (d) 180 (e) 181 (f) 182 (g) 183 and (h) 184 of $Zr_3Ir$.

The Fermi surface topology of $Zr_3Ir$ is quite complex; dominated by hole-like sheets. There are many segments of the Fermi surface close to each other. Therefore, multiband effects are expected to dominate the charge transport properties of $Zr_3Ir$.

### *3.8. Acoustic behavior and its anisotropy*

The relationship between sound velocity and thermal conductivity, $K = \frac{1}{3} C_v l v$ (where, $C_v$ is the specific heat per unit volume, $v$ is the velocity of sound in the solid, and $l$ is the mean free path for the lattice vibrations), makes sound velocity an essential characteristic of a material. In this section, we have calculated the sound velocities of the longitudinal and transverse modes in $Zr_3Ir$. The longitudinal ($v_l$) and the transverse ($v_t$) sound velocities can be calculated from the bulk ($B$) and shear ($G$) moduli using the following equations [80]:

$$v_t = \sqrt{\frac{G}{\rho}} \quad (45)$$

and

$$v_l = \sqrt{\frac{3B + 4G}{3\rho}} \quad (46)$$

Here, $\rho$ is the mass-density of the solid. According to these equations, a material's sound velocities are highly influenced by its density, and a material with zero shear modulus is unable to support the transverse mode of sound propagation.

The mean sound velocity, $v_m$, can be calculated using the longitudinal ($v_l$) and the transverse ($v_t$) sound velocities as follows [80]:

$$v_m = \left[\frac{1}{3}\left(\frac{2}{v_t^3} + \frac{1}{v_l^3}\right)\right]^{-\frac{1}{3}} \quad (47)$$

The acoustic impedance of a material controls how acoustic energy is transferred between two media, therefore, understanding it is useful.. A large sound pressure will produce a high particle velocity if the medium has a low impedance, but a high impedance will produce a relatively modest particle velocity from the same sound pressure. As a result, the majority of the sound is either transmitted or reflected depending on whether the impedance difference is roughly equal or significantly large. The study of acoustic impedance of materials with the same or different impedance from the surrounding medium has evolved over time into the development of instruments for noise reduction, transducer design, aircraft engine production, industrial factory design, and many undersea acoustic applications. The acoustic impedance of $Zr_3Ir$ was estimated using the equation given below [61,81]:

$$Z = \sqrt{\rho G} \quad (48)$$

where $G$ is the shear modulus and $\rho$ is the density of the material. The unit of acoustic impedance is the Rayl; 1 Rayl = $kgm^{-2}s^{-1}$ = 1 $Nsm^3$. A material having a high shear modulus and a high density has a high acoustic impedance.

A significant parameter in the construction of sound boards and loudspeakers is the intensity of sound radiation. A material's density and shear modulus are related to the intensity, *I*, of its acoustic radiation as [61,81]:

$$I \approx \sqrt{G/\rho^3} \qquad (49)$$

where, a high value of the combination of properties, $\sqrt{G/\rho^3}$ is defined as the *radiation factor*. The choice of materials for sound board design typically takes into account the radiation factor. Table 10 represents the calculated acoustic parameters.

**Table 10**

Calculated mass density ($\rho$ in gm cm$^{-3}$), longitudinal, transverse, and mean sound velocities ( $v_l$, $v_t$ and $v_m$ in ms$^{-1}$), acoustic impedance $Z$ (Rayl), and radiation factor $\sqrt{G/\rho^3}$ (m$^4$/kg.s of $Zr_3Ir$.

| Compound | *P* | $v_l$ | $v_t$ | $v_m$ | $Z$ ($10^6$) | $\sqrt{G/\rho^3}$ |
|---|---|---|---|---|---|---|
| $Zr_3Ir$ | 9.31 | 4209.68 | 1696.02 | 1920.75 | 15.79 | 0.18 |

The crystal structure and atomic arrangements have an impact on sound velocity (longitudinal and transverse). Each atom in a solid can vibrate in one of three modes (one longitudinal and two transverse modes). In anisotropic crystals, pure longitudinal and transverse modes are only supported along certain crystallographic directions. Pure transverse and longitudinal modes can be identified for [010], [100], [001], and [110] directions for crystals with tetragonal symmetry. Sound propagation modes in all other directions are found to be either quasi-transverse or quasi-longitudinal. The following relationships can be used to compute the acoustic velocities for tetragonal crystals along the principal directions [82]:

$$[100]\upsilon_l = [010]\upsilon_l = \sqrt{C_{11}/\rho};\ [001]\upsilon_{t1} = \sqrt{C_{44}/\rho};\ [010]\upsilon_{t2} = \sqrt{C_{66}/\rho}$$

$$[001]\upsilon_l = \sqrt{C_{33}/\rho};\ [100]\upsilon_{t1} = [010]\upsilon_{t2} = \sqrt{C_{66}/\rho}$$

$$[110]\upsilon_l = \sqrt{(C_{11} + C_{12} + 2C_{66})/2\rho};\ [001]\upsilon_{t2} = \sqrt{C_{44}/\rho};\ [1\bar{1}0]\upsilon_{t1} = \sqrt{C_{11} - C_{12}/2\rho}$$

where $\upsilon_{t1}$ and $\upsilon_{t2}$ refers to the velocity of first transverse mode and the second transverse mode, respectively, and $\upsilon_l$ is the velocity of the longitudinal mode. According to these relationships, a compound with a low density and high elastic constants will have high sound velocities. Directional sound velocities are displayed in Table 11.

**Table 11**

Anisotropic sound velocities ($ms^{-1}$) of $Zr_3Ir$ along different crystallographic directions.

| Propagation directions | | Anisotropic sound velocities | Ref. |
|---|---|---|---|
| [100] | [100]$\upsilon_l$ | 4159.03 | This work |
| | | 4229.00 | [3] Theo. |
| | [001]$\upsilon_{t1}$ | 1955.47 | This work |
| | | 1986.00 | [3] Theo. |
| | [010]$\upsilon_{t2}$ | 1521.42 | This work |
| | | 1567.00 | [3] Theo. |
| [001] | [001]$\upsilon_l$ | 4364.32 | This work |
| | | 4314.00 | [3] Theo. |
| | [100]$\upsilon_{t1}$ | 1521.42 | This work |
| | | 1567.00 | [3] Theo. |
| | [010]$\upsilon_{t2}$ | 1521.42 | This work |
| | | 1567.00 | [3] Theo. |
| [110] | [110]$\upsilon_l$ | 4280.64 | This work |
| | | 4332.00 | [3] Theo. |
| | [1$\bar{1}$0]$\upsilon_{t1}$ | 1135.08 | This work |
| | | 1256.00 | [3] Theo. |
| | [001]$\upsilon_{t2}$ | 1955.47 | This work |
| | | 1986.00 | [3] Theo. |

### *3.9. Thermophysical properties*

**(a) Debye temperature**

The Debye temperature ($\Theta_D$), one of the most important thermophysical parameters of solids, is related to a number of physical properties, including lattice vibration, thermal conductivity, phonon specific heat, interatomic bonding, resistivity, vacancy formation energy, melting temperature, and coefficient of thermal expansion. It also gives the characteristic energy of the phonons responsible for Cooper pairing in conventional superconductors. The Debye temperature is the temperature at which all the atomic modes of vibrations in a solid become active and it depends on the crystal stiffness and atomic masses of the constituent atoms. In general, solids having stronger interatomic bonding strength, higher melting temperature, greater hardness, higher acoustic wave velocity, and lower average atomic mass are found to have larger Debye temperatures. Moreover, Debye temperature marks the boundary between the classical and quantum behavior in lattice vibration. When $T > \Theta_D$, all the modes of vibrations roughly have an energy ~$k_BT$. On the other hand, when $T < \Theta_D$, the higher frequency modes are considered to be frozen and the quantum mechanical nature of vibrational modes is manifested [83]. In this study, $\Theta_D$ has been calculated from its proportionality to the mean sound velocity inside the crystal as follows [80,84]:

$$\Theta_D = \frac{h}{k_B}\left[\left(\frac{3n}{4\pi}\right)\frac{N_A\rho}{M}\right]^{\frac{1}{3}} v_m \qquad (50)$$

where $h$ is the Planck's constant, $k_B$ is the Boltzmann's constant, $n$ is the number of atoms in the unit cell, $M$ is the molar mass, $\rho$ is the density of the solid, $N_A$ is the Avogadro number, and $v_m$ denotes the mean sound velocity. The computed value of the Debye temperature was found to be 208.20 K, which is moderate and appropriate of solids with moderate hardness and melting temperatures.

**(b) Lattice thermal conductivity**

The lattice thermal conductivity of a solid is a useful parameter to investigate its applicability for high-temperature applications. The lattice thermal conductivity ($k_{ph}$) of a material at different temperatures determine the amount of heat energy carried by lattice vibration during thermal energy transport. For many technological applications, the lattice thermal conductivity is an important design parameter. For example, in thermoelectric (TE) conversions and in selecting thermal barrier coating (TBC) materials. For TBC one requires low value of $k_{ph}$, whereas high $k_{ph}$ materials are required as heat sinks in electrical devices to dissipate excessive thermal energy [85]. In a previous work, Slack developed the following empirical formula to estimate the $k_{ph}$ theoretically [86]:

$$k_{ph} = A\frac{M_{av}\Theta_D^3\delta}{\gamma^2 n^{2/3} T} \qquad (51)$$

In the above equation, $M_{av}$ is the average atomic mass in kg/mol, $\Theta_D$ is the Debye temperature in K, $\delta$ is the cubic root of average atomic volume in meter (m), $n$ is the number of atoms in the conventional unit cell, $T$ is the absolute temperature in K, and $\gamma$ is the acoustic Grüneisen parameter that measures the degree of anharmonicity of the phonons. A material with a low Grüneisen parameter value has low phonon anharmonicity, which leads to a high thermal conductivity. It is a dimensionless quantity that can be calculated from the Poisson's ratio using the following equation [86]:

$$\gamma = \frac{3(1 + v)}{2(2 - 3v)} \tag{52}$$

The factor $A(\gamma)$, due to Julian [87], is computed from:

$$A(\gamma) = \frac{5.720 \times 10^{7} \times 0.849}{2 \times (1 - {}^{0.514}/_{\gamma} + {}^{0.228}/_{\gamma^2})} \tag{53}$$

The estimated lattice thermal conductivity at room temperature (300 K) and the Grüneisen parameter are given in Table 12. The Grüneisen parameter of $Zr_3Ir$ is high; 2.615, implying that anharmonic effects are significant in the lattice dynamics of this compound. The phonon thermal conductivity at room temperature is also low.

**(c) Melting temperature**

Another important thermophysical parameter for solid is its melting temperature ( $T_m$) that gives the idea about the limit of temperature up to which it can be used. The melting temperature of a crystalline material is principally determined by its bonding energy and thermal expansion coefficient. A material with a high value of melting temperature shows strong atomic bonding, a high value of bonding energy and a low value of thermal expansion [88]. We have calculated the melting temperature ($T_m$) for $Zr_3Ir$ using the following equation based on the elastic constants [88]:

$$T_m = 354\,K + 4.5(K/GPa)\left(\frac{2C_{11} + C_{33}}{3}\right) \tag{54}$$

The calculated melting temperature of $Zr_3Ir$ is 1103.12 K. This value is also consistent with the moderate Debye temperature and hardness.

**(d) Thermal expansion coefficient and heat capacity**

The thermal expansion coefficient ($\alpha$) is an important thermomechanical parameter controlled by the bonding strength and lattice anharmonicity. Anharmonic vibrations leads to thermal expansion while the amplitude of these vibrations are dependent on crystal stiffness. This particular parameter is also closely linked to the melting temperature. The thermal expansion coefficient of $Zr_3Ir$ can be calculated using the following equation [81]:

$$\alpha = \frac{1.6 \times 10^{-3}}{G} \tag{55}$$

The thermal expansion coefficient is inversely linked to the melting temperature: $\alpha \approx 0.02/T_m$ [81,89]. The computed value of $\alpha$ of $Zr_3Ir$ is large as expected for systems with moderate melting point. The value of $\alpha$ is given in Table 12.

Another intrinsic thermal property of a material is its heat capacity. The change in the thermal energy per unit volume in a material per degree Kelvin change in temperature is defined as the heat capacity per unit volume ($\rho C_P$). Higher heat capacity materials have higher thermal conductivity and lower thermal diffusivity. Such materials have efficient thermal storage capacity. We have calculated $\rho C_P$ for $Zr_3Ir$ using the following equation [81]:

$$\rho C_P = \frac{3k_B}{\Omega} \tag{56}$$

where, $(1/_{\Omega})$ is the number of atoms per unit volume. The heat capacity per unit volume of $Zr_3Ir$ is shown in Table 12.

**(e) Dominant phonon mode**

Phonons play significant role in determining a variety of physical properties of condensed phases, including heat capacity, thermal conductivity, and electrical conductivity. The wavelength of the dominant phonon is the wavelength, $\lambda_{dom}$, at which the phonon distribution function attains a peak. The position of the peak is a function of the temperature and acoustic velocity. We have calculated $\lambda_{dom}$ at 300 K using following equation [89]:

$$\lambda_{dom} = \frac{12.566\, v_m}{T} \times 10^{-12} \tag{57}$$

The wavelength of the dominant phonon mode is longer in a compound with a greater average acoustic velocity, higher shear modulus, and lower density. The estimated value of $\lambda_{dom}$ in meter is presented in Table 12.

**Table 12**

Calculated Debye temperature $\Theta_D$ (K), lattice thermal conductivity $k_{ph}$ (W/m.K), melting temperature $T_m$ (K), thermal expansion coefficient α ($K^{-1}$), heat capacity per unit volume $\rho C_\rho$ ($JK^{-1}m^{-3}$), wavelength of the dominant phonon mode at 300 K $\lambda_{dom}$ (m), and Grüneisen parameter of $Zr_3Ir$.

| Compound | $\Theta_D$ | $k_{ph}$ | $T_m$ | α ($10^{-5}$) | $\rho C_\rho$ ($10^6$) | $\lambda_{dom}$ ($10^{-12}$) | $\gamma$ | Ref. |
|---|---|---|---|---|---|---|---|---|
| $Zr_3Ir$ | 208.20 | 0.40 | 1103.12 | 5.97 | 1.99 | 80.45 | 2.615 | This work |
| | 170.55 | --- | --- | --- | --- | --- | | [3]$^{Theo.}$ |

**(f) Minimum thermal conductivity**

At high temperatures greater than the Debye temperature, the phonon thermal conductivity of a compound becomes saturated to a limiting (minimum) value known as the minimum thermal conductivity ($k_{min}$). Actually, thermal conductivity achieves its lowest possible value at temperatures above the Debye temperature where the phonon wavelength becomes as small as the interatomic spacing. This particular thermal parameter is notable in that the existence of defects, such as dislocations, individual vacancies, and long-range strain fields associated with inclusions and dislocations, has no effect on it. This is mainly due to the fact that these defects affect phonon transport on length scales that are considerably larger than the interatomic spacing. Based on the quasi-harmonic Debye model, Clarke derived the following expression for calculating the minimum thermal conductivity, $k_{\min}$ of solids at high temperatures [89]:

$$k_{min} = k_B \upsilon_m (V_{atomic})^{-2/3} \quad (58)$$

where, $k_B$ denotes the Boltzmann constant, $\upsilon_m$ is the average sound velocity and $V_{atomic}$ signifies the cell volume per atom of the compound.

An elastically anisotropic solid also has anisotropic minimum thermal conductivity. The use of materials with anisotropic thermal conductivity is widespread, including heat spreading in electronic and optical device technologies as well as in heat shields, thermo-electrics, and thermal barrier coatings. The anisotropy in Minimum thermal conductivity is determined by the sound velocities in different crystallographic directions. Using the Cahill model [90], the following minimum thermal conductivities are determined along different directions:

$$k_{min} = \frac{k_B}{2.48} n^{2/3} (\upsilon_l + \upsilon_{t1} + \upsilon_{t2}) \quad (59)$$

The minimum thermal conductivities of $Zr_3Ir$ along the [100], [001], and [110] directions are shown in Table 13. For $Zr_3Ir$, the minimum thermal conductivities along different crystallographic axes are found to be greater than the isotropic minimum thermal conductivity.

**Table 13**

The number of atoms per mole of the compound $n$ ($m^{-3}$), minimum thermal conductivity (W/m.K) of $Zr_3Ir$ along different directions evaluated by the Cahill's method and Clarke's method.

| Compound | $n$ ($10^{28}$) | [001] $k_{min}$ | [110] $k_{min}$ | [100] $k_{min}$ | $k_{min}$ | Ref. |
|---|---|---|---|---|---|---|
| $Zr_3Ir$ | 4.81 | 0.55 | 0.54 | 0.56 | 0.35 | This work |
| | 4.79 | 0.55 | 0.56 | 0.57 | --- | [3]$^{Theo.}$ |

Table 13 also includes results from an earlier study [3]. The agreement between our computed values and the earlier ones are excellent. The minimum thermal conductivity of $Zr_3Ir$ is quite low. Low thermal conductivity solids with high damage tolerance are useful to be used as thermal barrier coating (TBC) material. In fact, the minimum thermal conductivity of intermetallic $Zr_3Ir$ is significantly lower than those of recently investigated MAX phase nanolaminates proposed as potential candidates for TBC applications [91-93].

### *3.10. Optical properties*

The interactions of electromagnetic radiation (photon) with the electrons in a crystalline solid are of decisive importance in optical modulation, optoelectronics, optical information processing and communications, display devices, and optical data storage. A crucial tool for analyzing the electronic energy band structure, DOS, state of impurity levels, formation of excitons, localized defect states, lattice vibrations, and certain magnetic excitations are also related to the frequency- or energy-dependent optical parameters, namely, the absorption coefficient $\alpha(\omega)$, optical conductivity $\sigma(\omega)$, dielectric function $\varepsilon(\omega)$, loss function $L(\omega)$, reflectivity $R(\omega)$, and refractive index $n(\omega)$ (where $\omega = 2\pi f$ is the angular frequency of the incident EMW) [94-97]. In this section, we have shown the results of the computed optical parameters of $Zr_3Ir$ for incident EMW energies up to 25 eV with the electric field polarizations along the [100] and [001] directions as depicted in Figs. 7(a-f). Electric field polarization direction dependent optical parameters illustrate the underlying optical anisotropy.

The absorption coefficient, $\alpha(\omega)$, specifies how far light of specific energy can penetrate into the material before being absorbed and helps us to understand the optimum solar energy conversion efficiency of a material. It also provides information on material's electrical nature, whether it is metallic, semiconducting, or insulating. As we can see from Fig. 7a, the absorption coefficient of $Zr_3Ir$ begins from 0 eV of photon energy for both polarization directions indicating its metallic electronic band structure. The absorption coefficient of $Zr_3Ir$ is quite high in the spectral region from ~ 2.60 to 20.0 eV, attaining a peak around 5.71 for [100] and 10.4 eV for [001] polarization directions in the ultraviolet (UV) region. The peak value for [100] is higher than that for [001] polarization direction indicating significant optical anisotropy in absorption characteristics.

The optical conductivity, $\sigma(\omega)$, is another important optoelectronic parameter of a material that can be defined as the conduction of free charge carriers over a defined range of the photon energies. For both polarization directions, photoconductivity starts with zero photon energy (Fig. 7b), which is a hallmark of metallic conductivity of $Zr_3Ir$, entirely consistent with the electronic band structure and TDOS calculations. The spectra of both polarizations exhibit notable variations in the range of photon energy from 0 to 5 eV and $Zr_3Ir$ has higher photo conductivity along [100] direction compared to [001] direction, demonstrating the anisotropic nature in optical properties. The optical conductivity decreases with increasing energy in the mid-UV region.

The dielectric function spectra have two parts; real and imaginary parts. The real part is connected to the electrical polarization of the material, while the imaginary part is connected to the dielectric loss. Fig. 7c shows that the real part of dielectric function crosses zero from below at ~ 18.10 eV, whereas the imaginary part flattens to a very low value at the same energy, indicating that the material should become transparent to the incident EMW above 18.10 eV. This phenomenon confirms the Drude-like behavior (metallic feature) of $Zr_3Ir$ under study. Both the real and imaginary parts of dielectric function show moderate optical anisotropy, particularly at low energies whereas at high energies of the electromagnetic wave, the anisotropy diminishes dramatically.

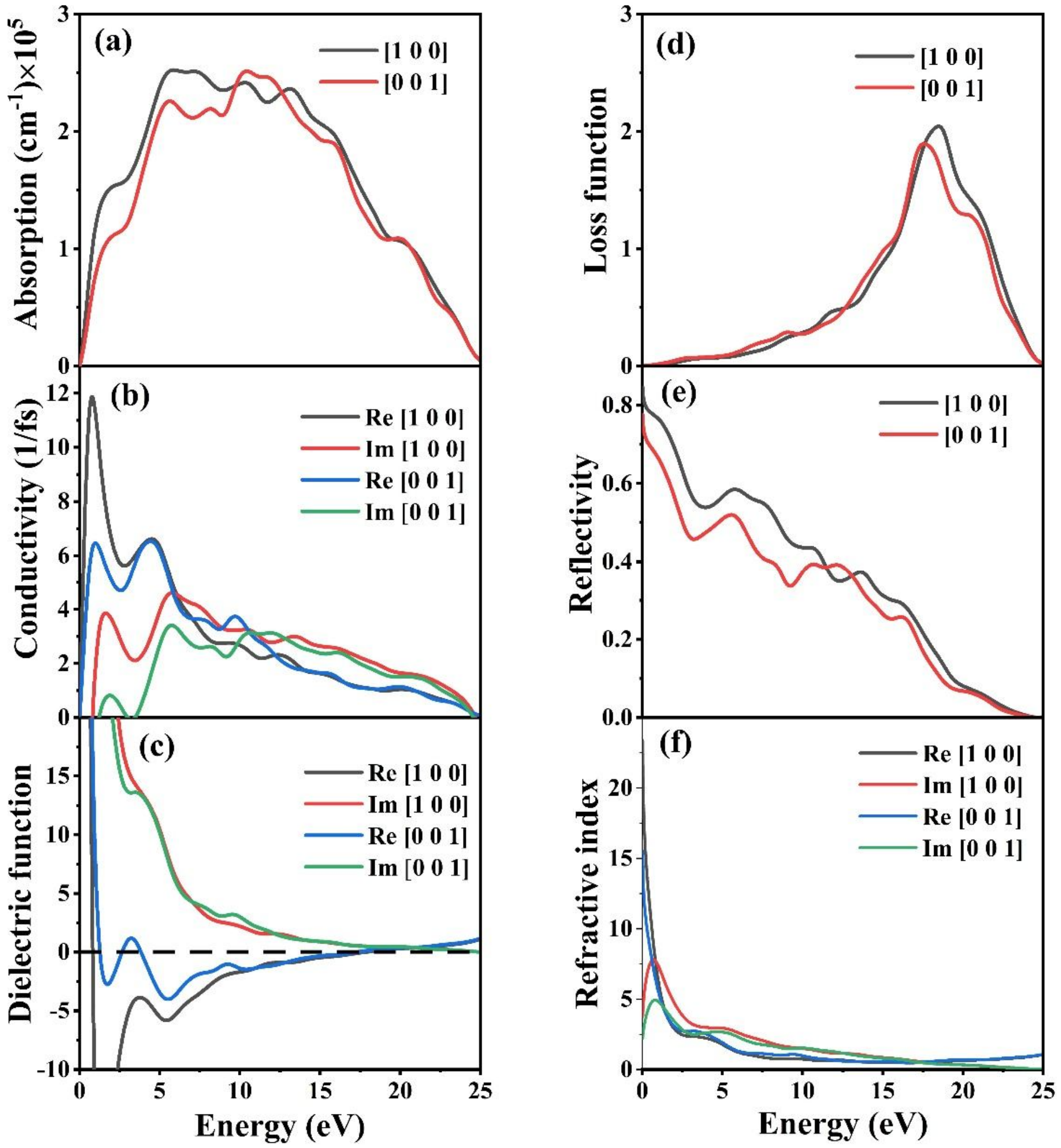

**Figure 7.** The frequency-dependent (a) absorption coefficient (b) optical conductivity (c) dielectric function (d) loss function (e) reflectivity, and (f) refractive index of $Zr_3Ir$ with electric field polarization vectors along [100] and [001] directions.

The loss function, $L(\omega)$, is an essential optical parameter describing the energy loss of a fast electron moving in a material. Under such a case, the fast-moving charge's energy is dampened, resulting in collective electronic excitations known as plasma excitations. When the energy matches the natural frequency, $\omega_P$, of the plasma excitation, the peak in the loss function emerges precisely. It is important to notice that the plasma energies coincide with the sharp falls in the absorption coefficient and reflectivity. Therefore, the compound under study is expected to behave transparently for photons with energies greater than the plasma energy and the optical properties should show behaviors similar to those of insulating systems. Fig. 7d shows that for [100] and [001] polarization directions, the peaks of $L(\omega)$ are located at 18.49 eV and 17.56 eV, respectively. Above the plasma energy, a metal's optical behavior shifts to dielectric-like response.

Reflectivity, $R(\omega)$, measures the fraction of the incident light energy reflected from the material. It is seen from Fig. 7e that $Zr_3Ir$ is a very good reflector of infrared and visible light. The highest reflectivity values of ~84% and ~76% are found at the infrared region for [100] and [001] polarization directions, respectively. The spectra have values above 50% and 45% from infrared to near-ultraviolet region for [100] and [001] polarization directions, respectively, and then start to decrease. As $R(\omega)$ falls sharply around 17 eV, $Zr_3Ir$ becomes transparent to the incident EMW having energies greater than 17 eV. Compounds having average reflectivity values more than 44% in the visible region have been found to reduce solar heating by reflecting a considerable percentage of the solar radiation spectrum [94-97]. So, the binary intermetallic $Zr_3Ir$ compound should be capable of reducing solar heating quite efficiently and can be used as a reflecting coating material.

The energy dependent refractive index, $N(\omega)$, is an optical parameter important for photonic device fabrication, such as for constructing optical wave-guides. In fact, $N(\omega) = n(\omega) + ik(\omega)$, where $k(\omega)$ is the extinction coefficient. The phase velocity of an electromagnetic wave in the material is determined by the real part of refractive index, whereas the imaginary part, often termed as the extinction coefficient, controls the attenuation of electromagnetic radiation inside the material. The $n(\omega)$ spectra (Fig. 7f) exhibit a slightly anisotropic nature up to 10.70 eV for both the polarizations. Above 10.71 eV of photon energy, no significant anisotropy is observed for $Zr_3Ir$. The calculated values of the static refractive index $n(0)$ are found to be at 18.90 and 14.21 along [100] and [001] directions, respectively. These values are quite high, and it is possible that the material under investigation might be used to improve the visual characteristics of electronic displays like LCDs and OLEDs. As expected, the peaks in $k(\omega)$ nearly match the spectrum of the imaginary part of the dielectric function.

All the optical spectra given in this section exhibit a moderate degree of anisotropies (comparatively higher at low energies) with respect to the polarization state of the incident EMW.

### *3.11. Superconducting state properties*

Previous studies have found superconductivity in $Zr_3Ir$ at 2.3 K [5]. Therefore, this compound is a low-$T_c$ superconductor. We have calculated $T_c$ of $Zr_3Ir$ using the widely used McMillan equation [98]:

$$T_c = \frac{\theta_D}{1.45} exp\left[-\frac{1.04(1 + \lambda_{ep})}{\lambda_{ep} - \mu^*(1 + 0.62\lambda_{ep})}\right] \qquad (60)$$

where $\mu^*$ is the Coulomb pseudopotential and $\lambda_{ep}$ is the electron-phonon coupling constant. $\mu^*$ is estimated before using Equation (44). The electron-phonon coupling constant can be estimated approximately using the widely used equation presented below [99]:

$$\lambda_{ep} = \frac{\gamma_{exp}}{\gamma_{cal}} - 1 \quad (61)$$

where $\gamma_{cal}$ and $\gamma_{exp}$ indicate the theoretical and experimental values of the electronic specific heat coefficient. The value of $\gamma_{exp}$ can be extracted from the experimental electronic heat capacity measurements as shown in Ref. [5] for $Zr_3Ir$. On the other hand, $\gamma_{cal}$ is found using the value of the DOS at the Fermi energy [100,101] as follows:

$$\gamma_{cal} = \frac{\pi^2 k_B^2 N(E_F)}{3}$$

The resulting $\gamma_{cal}$ value obtained using the $N(E_F)$ from the band structure calculations differs from the $\gamma_{exp}$ of $Zr_3Ir$. This is expected since the DOS value extracted from the band structure is not *dressed* properly by the electron-phonon interaction. The calculated value of $\lambda_{ep}$ using Equation 61 turns out to be 0.45. This value is somewhat lower than 0.56 obtained previously [5]. Using values of $\theta_D$, $\mu^*$ and $\lambda_{ep}$ obtained in our work, the superconducting transition temperature of $Zr_3Ir$, calculated using the McMillan equation is 1.9 K, quite close to the experimental value of 2.3 K [5]. The low value of $\lambda_{ep}$ suggests that $Zr_3Ir$ is a weakly coupled superconducting material. We predict that the matrix element of electron-phonon interaction potential, $V_{ep}$, is very low in $Zr_3Ir$. This follows from the relation $\lambda_{ep} = N(E_F)V_{ep}$ and the fact that the TDOS at the Fermi level is quite high [102,103]. The $N(E_F)$ is almost entirely due to the hybridized 4*d* and 4*p* electronic orbitals of the Zr atoms. We therefore infer that these electrons take part in the Cooper pairing in $Zr_3Ir$.

## 4. Conclusions

In this work, we have performed a detailed analysis of the structural, mechanical, electronic, optical, superconducting state and thermophysical properties of non-centrosymmetric superconductor $Zr_3Ir$ using a first-principles approach based on density functional theory. The majority of the findings dealing with the elastic, electrical, bonding, thermophysical, and optical properties are novel. The estimated lattice parameters and elastic constants correspond well with the previously published data [3–5].

From the results of formation enthalpy, mechanical parameters, and electronic structure, $Zr_3Ir$ is found to be stable. The compound is highly machinable and damage tolerant with ductile nature. $Zr_3Ir$ possesses moderate hardness and mixed bonding characteristics. The high value of Kleinman parameter of $Zr_3Ir$ indicates that the mechanical strength of $Zr_3Ir$ is mainly dominated by bond bending contribution. Based on the values of different elastic anisotropy factors ($A$, $A_1$, $A_2$, $A^U$, $A^{eq}$, etc.) and optical properties, we can say that $Zr_3Ir$ possesses a moderate level of elastic anisotropy. The electronic band structure, density of states, and Fermi surface reveal metallic characteristics of $Zr_3Ir$. There is significant electronic correlations in $Zr_3Ir$. The MPA and HPA predict that $Zr_3Ir$ has ionic and covalent bondings agreeing with the charge density map. The fracture toughness of $Zr_3Ir$ is high. The calculated Debye temperature is low for $Zr_3Ir$ indicating its moderate hardness and melting temperature. The minimum lattice thermal conductivity of $Zr_3Ir$ at high temperature is low, suggesting that this binary intermettalic compound has potential to be employed as a TBC material. The optical parameters show metallic features in agreement with the electronic band structure calculations. $Zr_3Ir$ shows very good reflectivity in the infrared and near visible regions that makes this compound capable of reducing solar heating and can be used as coating material. We have explored the superconducting state properties of $Zr_3Ir$ and found that it is a weakly coupled low-$T_c$ superconductor with low electron-phonon interaction energy. Since the phonon dynamics of this compound contains significant anharmonicity (noted from the high value of the Grüneisen parameter) and Fermi level is located close to the bonding peak in the DOS profile, application of pressure can change the superconducting transition temperature of $Zr_3Ir$.

**Acknowledgments**
This work was carried out with the aid of a grant (grant number: 21-378 RG/PHYS/AS_G - FR3240319526) from UNESCO-TWAS and the Swedish International Development Cooperation Agency (Sida). The views expressed herein do not necessarily represent those of UNESCO-TWAS, Sida or its Board of Governors.

**Declaration of interest**
The authors declare that they have no known competing financial interests or personal relationships that could have appeared to influence the work reported in this paper.

**Data availability**
The data sets generated and/or analyzed in this study are available from the corresponding author on reasonable request.

**CRediT author statement**

**Razu Ahmed:** Methodology, Software, Writing- Original draft. **Md. Sajidul Islam**: Software, Validation. **M.M. Hossain:** Writing- Reviewing and Editing. **M.A. Ali:** Writing- Reviewing and Editing. **M.M. Uddin:** Writing- Reviewing and Editing. **S.H. Naqib:** Conceptualization, Supervision, Formal analysis, Writing- Reviewing and Editing.